\documentclass[useAMS,usenatbib]{mn2e}

\usepackage[pdftex]{graphicx}
\usepackage{amsmath}
\usepackage{overpic}

\newcommand{\muu}{mag arcsec$^{-2}$}

\usepackage{relsize}

\newcommand{\hi}{%
  \relax
  \ifmmode
    \textrm{\textsc{HI}}
  \else
    \textsc{H{\smaller}I}
  \fi
}

\newcommand{\simgt}{\lower.5ex\hbox{$\; \buildrel > \over \sim \;$}}
\newcommand{\simlt}{\lower.5ex\hbox{$\; \buildrel < \over \sim \;$}}

   \title[Stellar disc truncations]{Stellar Disc Truncations and Extended Haloes in Face-on Spiral Galaxies}
\author[S. P. C. Peters et al.]{S. P. C. Peters$^{1}$,
P. C. van der Kruit$^{1}$\thanks{For more information, please contact P.C. van der Kruit by email at vdkruit@astro.rug.nl.}, 
J.~H. Knapen$^{2,3}$, I. Trujillo$^{2,3}$, J. Fliri$^{2,3}$,\\
\ \\
{\LARGE {\rm M. Cisternas$^{2}$, L. S. Kelvin$^{2,3,4}$}}\\
$^{1}$Kapteyn Astronomical Institute, University of Groningen, P.O.Box 800, 9700AV Groningen, the Netherlands\\
$^{2}$Instituto de Astrof\'{\i}sica de Canarias, VíaL\'actea, E-38205 La Laguna, Tenerife, Spain\\
$^{3}$Departamento de Astrof\'{\i}sica, Universidad de La Laguna, E-38205 La Laguna, Tenerife, Spain\\
$^{4}$Institut f\"ur Astro- und Teilchenphysik, Universit\"at Innsbruck, Technikerstra\ss e 25' 6020 Innsbruck, Austria\\
}

\begin{document}
\date{Accepted 2015 month xx. Received 2015 Month xx; in original form 2015 Month xx}
\pagerange{\pageref{firstpage}--\pageref{lastpage}} \pubyear{2015}

\maketitle

\label{firstpage}

  \begin{abstract}
We use data from the IAC Stripe82 Legacy Project to study the surface 
photometry of 22 
            nearby, face-on to moderately inclined spiral galaxies. 
The reprocessed and combined Stripe 82 $g'$, $r'$ and $i'$ images 
            allow us to probe the galaxy down to 29-30 
$r'$-magnitudes/arcsec$^2$ and thus 
            reach into the very faint outskirts of the galaxies. 
            Truncations are found in three galaxies.
            An additional 15 galaxies are found to have an apparent
extended stellar halo. 
Simulations show that the scattering of light from the inner galaxy 
by the Point Spread Function (PSF) can produce faint structures resembling 
haloes, but this effect is insufficient to 
fully explain the observed haloes.
            The presence of these haloes and of truncations is mutually 
exclusive, and we argue that the presence of a stellar halo and/or light 
scattered by the PSF can hide truncations.
            Furthermore, we find that the onset of the stellar
halo and the truncations scales tightly with galaxy size.
            Interestingly, the fraction of light does not correlate with 
dynamic mass.  Nineteen galaxies are found to have breaks in their 
profiles, the radius of which also correlates with galaxy size.
\end{abstract}

\begin{keywords}
galaxies: photometry, galaxies: spiral, galaxies: structure
\end{keywords}

\section{Introduction}
Truncations in edge-on stellar discs were first discovered by \citet{vdk79}, 
who noted that the radial extent of these galaxies did not grow with deeper 
photographic exposures, in contrast with the vertical extent.
Truncations are very sharp, with scalelengths of less than 1 kpc and typically 
occur at around four to five times the exponential scalelength of the inner 
disc \citep{vdks81a,vdks81b,vdks82,fry98,djetal07b,Barteldrees1994,kkg02}.
In a recent extensive analysis of 70 edge-on galaxies in the near-IR, taken 
from the {\it Spitzer Survey of Stellar Structure in Galaxies} (S$^4$G), 
\citet{Com12} concluded that about three out of four thin discs are
truncated. 

While truncations in edge-on galaxies have been observed for the last 35 
years, their face-on counterparts remain elusive.
This is in part due to shorter line-of-sight integration through face-on 
and moderately inclined galaxies.
Due to this, the expected surface brightness at four scalelengths is about 
26 $B$-mag arcsec$^2$, only a few percent of the dark 
sky \citep{vdks81a,vdks82}.
In recent times, various studies on inclined or face-on samples have been 
performed looking for such truncations. For comparison 
with the van der Kruit \&\ Searle systems one would first 
concentrate on non-barred late-type systems.
\citet{pt06} studied such
a sample of 90 moderately inclined, late-type systems 
through ellipse-fitting of isophotes in Sloan Digital Sky Survey (SDSS) 
data. 
They identified galaxies with truncations, but they also found
cases where the discs continued as unbroken exponentials, or even showed an 
upturn in the outer parts. This latter behaviour was earlier found in barred 
early-type galaxies by \citet{Erwin2005}, followed up by 
\citet{Erwin2008A}  on 66 barred and by \citet{GEAB11} on 47 
unbarred, early-type galaxies. 
These studies led to a classification system, in which Type I designated 
exponentials out to the noise level, Type II with a turndown of the profile
before that and Type III with an upturn.

\citet{pt06} found that 60\%\ of their galaxies had a break in the 
profile between 1.5-4.5 times the scalelength followed by a steeper
downbending. In 26 of their galaxies they included in their classification (in 
addition to the class Type II) a designation CT for `classical truncation'.
These, they proposed where similar features as the truncations in edge-on 
systems.

This result has been disputed by \citet{vanderKruit2008A}, who argued 
that these are in fact breaks similar to those found by \citet{Freeman1970A}.
Originally,
\citet{vdks81a,vdks82} had found that the truncation in edge-on systems 
occurred at $4.2\pm0.5$ scale lengths, while \citet{kk04} quoted $3.6\pm0.6$ 
scale lengths. In the latter paper it was estimated that this corresponded 
to a face-on surface brightness at the onset of the truncations of $25.3\pm0.6$ 
R-mag arcsec$^{-2}$. In contrast the radii of the features in the \citet{pt06}
study were at $2.5\pm0.6$ scale lengths and surface brightness 23 to 25 in 
$r^{\prime}$ (see their Figure 9).
The view of breaks and truncations as two separate features was 
proposed also by \citet{mbt12}.
In a study of 34 edge-on spiral galaxies, they found that the innermost 
break occurs at $\sim\!8\pm1$ kpc and truncations at $\sim\!14\pm2$ kpc 
in galaxies.

Another reason for the elusiveness of face-on truncations is the lopsided 
nature of spiral galaxies at faint levels.
This is clearly demonstrated by the $m=1$ Fourier component of surface 
brightness maps \citep{rz95,zr97,zr13} and in polar projected contour maps 
\citep{pdla02}.
In face-on galaxies like NGC 628 an isophotal map shows that 
the outer contours have a much smaller spacing than the inner ones 
\citep{shvdk84,vdk88}.
Due to the lopsided nature of these galaxies, the typical method of 
fitting ellipses to the profile will smooth out the truncations.
\citet{pkj13} devised two alternative methods to derive the profiles.
Although the techniques looked promising, they were hampered by the 
limited sky brightness of the SDSS images and did not detect any truncations.

A third reason may be that at the expected surface brightness levels in
face-on galaxies, truncations will be of similar brightness as stellar haloes.
In $\Lambda$CDM cosmology, galaxies build up hierarchically, with small 
objects merging together under their mutual gravity to form ever larger 
objects, leading up to the structures we see today \citep{White1978}. 
The lopsided outer edges of spiral galaxies are caused by tidal 
interactions and in-fall from these structures.
Stellar haloes are remnants of this merger process, a graveyard of 
long-gone galaxies.
Faint stellar haloes have been successfully detected around 
spiral galaxies. 
This is often done by resolving the individual stars, using very deep 
observations, for example as those obtained using the {Isaac Newton 
Telescope} by \citet{Ferguson2002}, or the {Hubble Space Telescope} 
by \citet{Monachesi2013}. 
Using surface photometry, \citet{bt13} found that at very faint 
levels the radial profiles of seven disc galaxies indicate the 
presence of a stellar halo at levels of 29 to 30 $r-$mag/arcsec$^2$, 
but found no evidence for truncations. 
They argue that these haloes may be the cause of the hidden truncations 
in the radial profiles of face-on galaxies.
Recently, \citet{mtk14} used a theoretical model to demonstrate that a 
stellar halo can indeed outshine truncations in a face-on galaxy.

In this context it is relevant to consider the up-bending profiles, 
designated Type III. As noted above this behaviour was first found by 
\citet{Erwin2005}; \citet{pt06} found this in 30\%\ of their galaxies.
In the latter paper a prominent example of this was NGC\,3310, which is
a well-known case of an outer disturbance. In this system it is in the form of 
a `bow and arrow', which would be the cause of the upbending profile. 
NGC\,3310 has been studied in detail by e.g. \citet{vdk76} 
and \citet{ks2001} and is a clear case of a merger. This led \citet{kf11}
to examine deep SDSS images of the Type III galaxies
in \citet{pt06}, which indicated that many, if not all, 
of these show signs of interaction or distortion at faint levels. 

The presence of Type III profiles has been established in galaxy samples
covering a wide range of Hubble types and environments 
\citep{Roed12, Herr13, Laine14}. These may represent outer disc structures
not related to merging and that are intrinsic to the disc structure. To
ensure that we do not include any distorted outer discs, we have in the present 
study concentrated on undisturbed systems, selected on deep SDSS images, 
and such systems would be excluded.

The aim of this paper is to detect truncations and stellar haloes in 
surface photometry using data for a sample of face-on and inclined 
spiral galaxies available in the new IAC Stripe 82 Legacy Project 
\citep[in preparation]{ft13}.
This project is a careful reprocessing of the SDSS Stripe 82 data, 
improving over the regular SDSS data by $\sim\!2$ magnitudes.
This allows us to trace the profiles down to $\sim\!30$ $r-$mag/arcsec$^2$, 
a depth which few studies have reached before 
\citep[e.g.][]{Zibetti2004A, Zibetti2004B, Jablonka2010, bt13, tb13}
We use four different photometric techniques to analyse 
the data, the classical ellipse fitting routines, the Principle Axis Summation (PAS) and Equivalent Profiles (EP) from \citet{pkj13} and 
a new technique that we call Rectified Polar Profiles, introduced here.
We find that using these techniques, we can indeed identify 
both stellar haloes and truncations in galaxies.

The remainder of this paper has the following outline. 
In Section \ref{sec:samples} we describe the sample and the data reduction. 
Section \ref{sec:photometric} introduces the various photometric 
techniques we used. 
The results are presented in Section \ref{sec:results}, followed 
by a discussion in Section \ref{sec:discussion}. We summarize the 
results in Section \ref{sec:conclusions}. The results for the 
individual galaxies are presented in the online Appendix.

\section{Galaxy Sample}\label{sec:samples}
\subsection{Sample Selection}
The goal of our project is to derive stable, extremely deep surface photometry 
of face-on to moderately inclined disc galaxies.
Our initial sample consisted of all galaxies in the SDSS Stripe 82 sample 
with a diameter larger than one arc-minute.
This set consists of  177 elliptical, irregular, face-on and edge-on disc 
galaxies.

\begin{figure}
\centering
\includegraphics[width=0.48\textwidth]{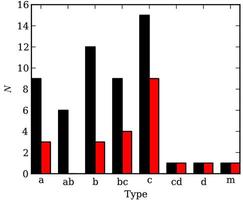}
\caption[Histogram of the distribution of Hubble types for the sample]{Histogram of the distribution of Hubble types for the sample of disc galaxies. The left black bars shows the distribution for the 54 initial disc galaxies that were retrieved. The right red bars show the distribution of the final 22 galaxies used in our analysis. Classification based on the primary listing in NED.}\label{fig:histogramsample}
\end{figure}

The SDSS jpeg colour images for each galaxy were downloaded using the Finding 
Chart Tool\footnote{The URL is cas.sdss.org/dr7/en/tools/chart/chart.asp.}.
Each image was visually inspected for type, possible signs of distortions, 
mergers and nearby objects, which might cause problems with the photometry.
We accepted those face-on to moderately inclined disc galaxies for which 
we found no potential issues. The criterion was then a judgement that 
the discs were visible sufficiently well to derive radial profiles.
In total these were 54 galaxies. 
These galaxies were then retrieved from the IAC Stripe 82 Legacy Project 
pipeline.
The images were again visually inspected for any form of distortion, 
nearby objects and possible issues of the sky-levels.
This left us with a final sample of 22 galaxies, for which we are 
confident of the extracted surface photometry down to 29-30 $r'$-mag/arcsec$^2$.
The primary properties of this sample are shown in Table \ref{tbl:sample}.
We show the distribution of Hubble types in Figure \ref{fig:histogramsample}.

\begin{table*}
\centering
\begin{tabular}{l|cccccccccc}
Galaxy         & Type & $M_\textrm{b}$& $M_\textrm{B}$& $v_\textrm{rad}$ [km/s]& $D$ [Mpc]& $i$ [$^\circ$] &  $V_{\rm rot}$  [km/s] & PA [$\theta$]\\
\hline \hline
IC\,1515 	&	 Sb 	&	14.4	&	-20.5	&	6560	&	93.1	&	26.8	&	332.6	&	252.3		\\
IC\,1516 	&	 Sbc 	&	14.2	&	-21.0	&	7170	&	101.7	&	17.6	&	418.7&	202.1			\\
NGC\,450 	&	 SABc 	&	12.8	&	-19.3	&	1729	&	24.5	&	45.5	&	146.4&	171.6			\\
NGC\,493 	&	 Sc 	&	13	&	-19.7	&	2277	&	32.3	&	72.9	&	158.2&	148.9	\\
NGC\,497 	&	 SBbc 	&	13.8	&	-21.6	&	7932	&	112.5	&	65.4	&	302.9&	219.5			\\
NGC\,799 	&	 SBa 	&	14.2	&	-20.5	&	5730	&	81.3	&	28.9	&	341.8&	159.1			\\
NGC\,856 	&	 Sa 	&	14.2	&	-20.5	&	5786	&	82.1	&	38	&	284.2&	113.6			\\
NGC\,941 	&	 SABc 	&	13	&	-18.8	&	1557	&	22.1	&	44.7	&	123.0&	254.1			\\
NGC\,1090 	&	 Sbc 	&	12.6	&	-20.4	&	2660	&	37.7	&	64.4	&	188.2&	190.6			\\
NGC\,7398 	&	 Sa 	&	14.8	&	-19.6	&	4735	&	67.2	&	48.4	&	235.4&	163.1			\\
UGC\,139 	&	 SABc 	&	14.5	&	-19.3	&	3899	&	55.3	&	64.6	&	172.7&	170.6			\\
UGC\,272 	&	 SABc 	&	15.2	&	-18.5	&	3821	&	54.2	&	67.2	&	133.8&	219.0			\\
UGC\,651 	&	 Sc 	&	16.1	&	-18.4	&	5064	&	71.8	&	68.9	&		&	256.3		\\
UGC\,866 	&	 Sd 	&	15.6	&	-16.5	&	1710	&	24.3	&	71.2	&	80.3&	146.9			\\
UGC\,1934 	&	 Sbc 	&	15.5	&	-20.8	&	11996	&	170.2	&	74.7	&	297.5	&	200.4		\\
UGC\,2081 	&	 SABc 	&	15	&	-17.8	&	2522	&	35.8	&	54.7	&	122.0&	163.1			\\
UGC\,2311 	&	 Sb 	&	14.1	&	-21.1	&	6901	&	97.9	&	30.9	&	343.0	&	170.2		\\
UGC\,2319 	&	 Sc 	&	15.5	&	-19.7	&	6827	&	96.8	&	72.8	&	202.6&	45.5			\\
UGC\,2418 	&	 SABb 	&	15.1	&	-20.2	&	6671	&	94.6	&	46.7	&	205.4&	184.5			\\
UGC\,12183 	&	 Sbc 	&	15.5	&	-19.0	&	4728	&	67.1	&	54.8	&	164.6&	120.4			\\
UGC\,12208 	&	 Scd 	&	15.2	&	-18.7	&	3342	&	47.4	&	42.8	&		&	156.2		\\
UGC\,12709 	&	 SABm 	&	14.4	&	-18.6	&	2700	&	38.3	&	51.8	&	103.7	&	229.7		\\
\hline\hline
\end{tabular}
\caption[Sample properties]{
Sample properties. Morphological type, apparent magnitude $M_\textrm{b}$, absolute magnitude $M_\textrm{B}$ (both in $B$-band), 
radial velocity $v_\textrm{rad}$ and maximum rotational velocity $V_{\rm rot}$ from HYPERLeda, based on width at $20$ of the HI flux and corrected based on the inclination from this work \citep{HYPERLeda,Springob05,Theureau98,Meyer04}. 
Distance $D$ from NED, based on Virgo + GA + Shapley and assuming $H_{0}$ 70.5 km/sec$\cdot$Mpc, $\Omega_{\rm matter}$ 
0.27 and $\Omega_{\rm vacuum}$ 0.73. The absolute magnitude has been distance corrected using this distance and has been corrected for foreground extinction based on \citet{Schlafly11}. Inclination $i$ this work.
No rotational velocities are available in the literature for UGC\,651 and UGC\,12208. Position Angle PA based on this work.
}\label{tbl:sample}
\end{table*}

\subsection{Image Reduction and Calibration}

The IAC Stripe 82 Legacy Project \citep{ft13}
produces (co-)adds of the whole Stripe 82 data set using an 
automated (co-)addition pipeline. 
The pipeline queries all available
data from the SDSS archive, aligns them photometrically and calibrates
them to a common zero-point using calibration tables provided by SDSS
for each image. 
Accurate values of the sky background are determined
on object-masked images and subtracted from the calibrated images.
Images with large sky backgrounds, notable sky gradients, large PSF
widths and images affected by clouds or taken under bad transparency
conditions are removed from the final stack.  
Finally, all images
passing the selection criteria are re-gridded onto a field of view of
0.25 square degrees and median-combined by SWarp \citep{bertin02}.
The IAC Stripe 82 Legacy Project is presented at 
www.iac.es/proyecto/stripe82. The (co-)added images produced for
this project are considerably smaller in size then in the data release
(100 compared to 900
square arcminutes), and the images have been centred on
each of the target galaxies to facilitate profile extraction.

The (co-)added images are calibrated to a zero-point $aa\!=\!-24$\,mag,
and are already corrected for airmass and extinction. With an area
$dS=0.396''\!\times\!0.396''$ of each pixel in SDSS and an exposure
time of 53.907456\,s, the conversion from counts to magnitudes can
be performed by using Equation \ref{eqn:calibration}:
\begin{eqnarray}
 \mu_0 &=& aa + 2.5\log_{10}\left(\mathrm{exptime} \times dS\right) \label{eqn:ZP}\,\,,\\
 \mathrm{mag} &=& -2.5 \log_{10}\left( \mathrm{counts}\right) + \mu_0                   \,\,.  \label{eqn:calibration} 
\end{eqnarray}
Since the mosaicking process leaves the various bands at exactly the
same pixel position, combining the $g'$, $r'$ and $i'$ images is
simply a matter of taking the average of the three.  
We do this to increase the signal-to-noise ratio and thus to extend 
our analysis further out in the galaxies.
Combining the three
bands is destructive to the zero-point calibration process and we thus need to
re-calibrate the data.  We measure ellipse profiles on both the original $r'$
and new combined images.  We find the new calibration zero-point
$\mu_0$ of the (co-)added profiles by shifting the profiles up or down,
until they align with the calibrated $r'$-band profiles.

\subsection{Object Masking and Background Treatment}
\begin{figure*}
\centering
\includegraphics[width=0.48\textwidth]{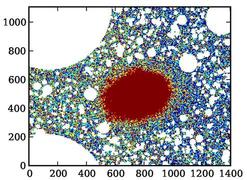}
\includegraphics[width=0.48\textwidth]{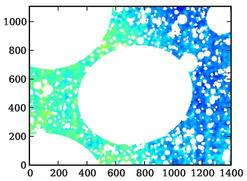}
\includegraphics[width=0.48\textwidth]{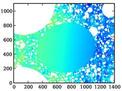}
\includegraphics[width=0.48\textwidth]{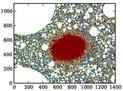}
\caption[Overview of the background modelling process]{Overview of the background modelling process for the galaxy {NGC\,941}. Top left: the original image with objects masked. Top right: the model created for the sky, excluding the galaxy. Lower left: the sky model including the galaxy. Lower right: the final image with the sky model subtracted from the observation. In all figures, the colour scales run from -1.0 to 1.0 ADU. Horizontal and vertical axes are in pixels (one pixel = $0.396''$). North is right, east is top.}
\label{fig:backgrounds}
\end{figure*}

Any study of surface photometry requires a careful removal of background and 
foreground objects. 
At the 29-30 $r'$-mag/arcsec$^2$ depth we aim to reach in the case of the IAC 
Stripe 82 Legacy Project, bright stars far outside the frame can still 
contribute flux.
We therefore need to model and remove this contamination.

As a first step, a smoothed and stacked version of the various bands was 
inspected and object masks were defined by hand.
By repeating this process multiple times, we were able to  remove all 
faint background sources from the image.
At this stage, an elliptical mask was defined for which we were certain 
that it comfortably covered the entire galaxy. 
The remaining background was then modelled using the following method. 
Around each pixel that was not part of an object mask and not in the 
galaxy mask, a $21\times21$ pixel ($\sim8.3''\times8.3''$)  box was 
defined, from which we selected a sample of all pixels not part of any mask.
A sigma-clip of $2.5\sigma$ was run over this sample, after which the 
background value was measured as the median of the remaining pixels.
By repeating this process for each pixel, a map of the large-scale 
structure in the background was created.
In this background model, we then selected a 50 pixel wide ring around 
each galaxy mask, to which we fit a 2D linear plane. This plane was 
extrapolated to model the background inside the galaxy mask.
The complete background model was subsequently subtracted from the original 
image.
We demonstrate the background subtraction process in 
Figure \ref{fig:backgrounds}.\label{sec:background}

Our next step was to measure the uncertainty and noise {in the background}.
This was done on the original combined data, using the same mask as the one 
used in the background-modelling step.
There were always over 200,000 pixels used in the measurement.
We use the Python package \textsc{scipy.stats.bayes\_mvs} to perform a 
Bayesian estimation of the confidence in the mean and the standard 
deviation \citep{o06}.
The uncertainty is based on the average confidence limit for the mean.
The pixel-to-pixel noise is taken as the standard deviation of the sample.
This is indicated in the figures in the online appendix.

\section{Profile Extraction}\label{sec:photometric}
In this section we discuss our four methods to extract radial surface brightness
profiles from the observed distributions. The first three have been used 
before, but the fourth one is new.

\subsection{Ellipse Fitting}\label{sec:ellipse}
We use the IRAF package \emph{stsdas.ellipse} \citep{j87,Busko1996} to extract 
ellipse profiles from the data.
In a first pass, the package was allowed to run with the position angle and 
inclination as free parametres. 
This was done to get the best estimates for the various parameters.
We use the innermost ellipse for the exact position of the center of the galaxy.
The position angle and inclination were measured from the 25th magnitude 
ellipse, where we define the inclination as $i=\arccos(b/a)$, using major 
and minor axis $a$ and $b$. This  also gives us the ellipticity $e$.
In the second pass, the central position, position angle and inclination 
were fixed and the method was run again to produce the final radial profiles.

\subsection{Equivalent Profiles}
The original method for deriving the photometry of galaxies was the use of 
the Equivalent Profiles method EP \citep{Vaucouleurs1948A}.
The modern-day version of the EP was introduced by \citet[]{pkj13}, 
in an attempt to understand the reliability of the ellipse fitting.
Compared to the latter, EP works in a fundamentally different way.
Rather than taking a radius $R$ and measuring a corresponding average 
surface brightness $\mu(R)$, this method works in the opposite direction.
Starting  with the lowest magnitude in the image and working up from there, 
for each magnitude it counts the number of pixels $N$ that are equal to or 
brighter than the current magnitude.
We make the reasonable assumption that the galaxy forms an ellipse on the 
sky, with a shape based on the 25th magnitude parameters from the ellipse 
fitting in Section \ref{sec:ellipse}.
Since each pixel covers a small surface on the sky of $0.396''\times0.396''$, 
we can use the total amount of pixels $N$ belonging to brightness $\mu$ to 
estimate a total surface $S$ and from there use the ellipse to estimate an 
equivalent (major axis) radius $R$.

\citet{pkj13} extensively tested the technique on a sub-sample of 29 galaxies 
defined previously by \citet{pt06}, and found that the results produced by 
EP and ellipse fitting are remarkably similar. 
At magnitudes near the pixel-to-pixel noise, EP tends to go wrong as it 
includes noise peaks into the equivalent radius. 
\citet{pkj13} set out to include adaptive smoothing and masking of the 
background to avoid this problem, but overall the method was found to 
work reliably only above the pixel-to-pixel noise limit. 

In this work, we rotate and centre the original image, based on the 
positioning as measured using the ellipse profiles in 
Section \ref{sec:ellipse}. 
We define an elliptical mask with a trust radius $R_\mathrm{trust}$ set by 
eye, beyond which the data are set to blank.
Similar to \citet{pkj13}, we again plot our profiles out to the radius 
where over- or under-subtracting the data by two times the uncertainty 
leads to a deviation of more than 0.2 magnitudes, compared to the original 
profile.

\subsection{Principle Axis Summation}
In an attempt to  recognize the truncations seen in edge-on galaxies with 
respect to face-on galaxies better, \citet{pkj13} developed the Principle
Axis Summation method PAS.
The PAS  has been developed to project face-on galaxies into edge-on 
galaxies, in the hope that features similar to truncations appear.
While the method itself was successful, true truncations remained elusive, 
due to the much 
shallower nature of the original SDSS data \citep{pkj13}.

Similar to the EP, we again rotate and centre the image, after which we 
apply the same elliptical blanking mask at trust radius $R_\mathrm{trust}$.
The PAS  then partitions the galaxy into four quadrants.
Quadrants with visible problems, for example a foreground star, were removed.
Each quadrant is then summed onto the major axis, after which the main 
profile is formed by the median of the accepted quadrants.
A dynamic binning algorithm is applied to ensure that the profile has 
a signal-to-noise ratio of at least two. 

The uncertainty in each point is different, due to the varying number 
of pixels involved in the sum. 
Similar to \citet[]{pkj13}, 
for each point, we repeat the PAS technique many times on randomly drawn 
samples from the set of background pixels defined in 
Section \ref{sec:background}, which gives a good estimate of the uncertainty.

\begin{figure}
 \centering
  \includegraphics[width=0.48\textwidth]{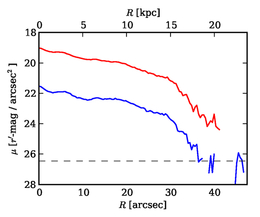}
  \caption[A truncation in an edge-on galaxy seen using PAS]{A truncation in the edge-on galaxy {UGC\,2584}. The lower, blue curve shows the $r'$-band surface brightness measured in a strip along the central plane of the galaxy. The top, red curve shows the PAS profile for this galaxy in arbitrary units. As can be seen clearly from the figure, the truncation is visible in both methods, starting at $R\simeq30''$ and 15\,kpc. The dashed, gray line represent the pixel-to-pixel noise.}
  \label{fig:UGC02584}
\end{figure}
\label{sec:PAStech}

To demonstrate the effect of PAS and the way we expect truncations to 
appear in face-on galaxies, we have applied the method to the edge-on 
galaxy {UGC\,2584}, for which the data are also available in the IAC 
Stripe 82 Legacy Project, and which we have reduced in the exact same 
ways as the face-on galaxies in this sample.
The PAS results are shown as the red (top) curve in Figure 
\ref{fig:UGC02584}, whereas the blue (lower) curve demonstrates 
the surface brightness measured along the central plane of the galaxy.
It is clear that a truncation is visible at a radius of 15 kpc, 
evident from both methods.

As noted in \cite{pkj13}, the PAS profiles have units of magnitudes/arcsec, 
as the minor axis summation can reach higher values in nearby galaxy with 
large apparent sizes. 
The result of this is that while the profile itself is reliable, its 
vertical offset and extent compared to the other profiles varies.
For the sake of comparison, we therefore choose to fix the central 
brightness as seen from the PAS at one magnitude lower than the central 
brightness as measured from the ellipse profiles.

\begin{figure*}
 \centering
\includegraphics[width=0.48\textwidth]{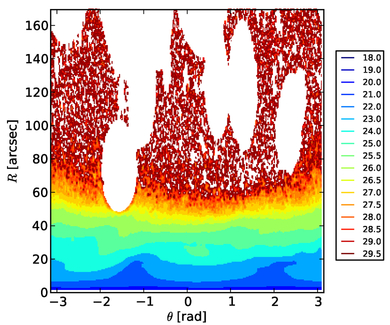}
\includegraphics[width=0.48\textwidth]{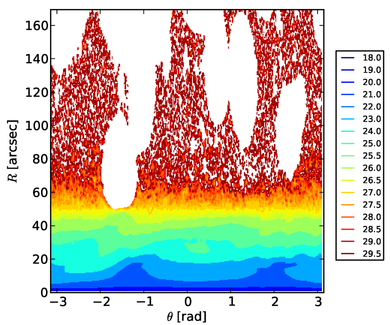}
\includegraphics[width=0.48\textwidth]{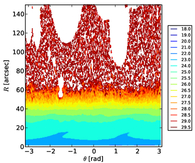}
\includegraphics[width=0.48\textwidth]{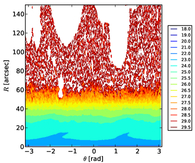}
\caption[Examples of rectified polar plots]{Polar plots for two of our galaxies, IC\,1515 (upper) and UGC\,12208 (lower). On the left the polar plot before rectification and on the right after rectification according to the procedure described in section \ref{sec:PolarPlots} at a surface brightness of 26 mag/arcsec${^2}$. The minor axis is at position angles 0 and 
$\pm \pi$. }
  \label{fig:PolarPlots}
\end{figure*}

\begin{figure*}
 \centering
\includegraphics[width=0.48\textwidth]{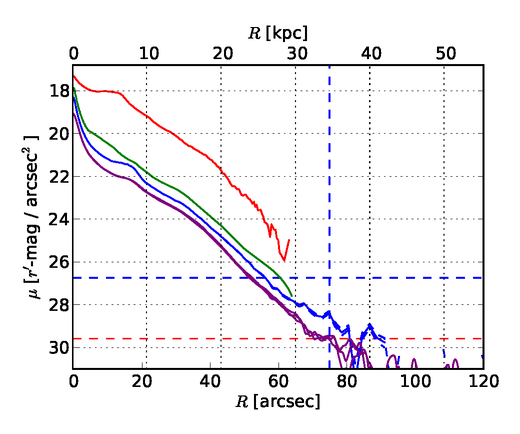}
\includegraphics[width=0.48\textwidth]{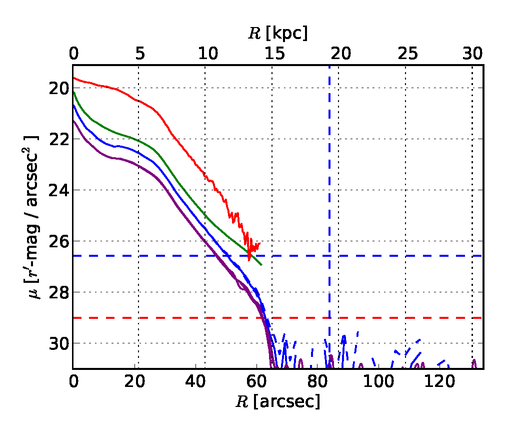}
\caption[Examples of radial profiles]{Full set of profiles for two examples galaxies, IC\,1515 and UGC\,12208. The red upper profile is from the PAS and the green is from the EP. The blue lines are the conventional ellipse fitting profiles. These are offset by +0.5 mag. The purple profiles show the results of the RPP applied at surface brightnesses of 25, 26 and 27 mag/arcsec$^{2}$; these are offset by +1.0 mag. The dashed blue vertical lines denote the maximum radius used for the PAS and EP profile extraction. The dashed horizontal blue lines represent the $1\sigma$ standard deviation in the pixel-to-pixel noise and the red dashed horizontal lines the uncertainty limit. }
  \label{fig:Combined}
\end{figure*}

\begin{table}
\centering
\begin{tabular}{l|cc|cc|cc|c}
Galaxy	&	x1	&	x2	&	x3	&	x4	&	x5	&	x6	& $r_\mathrm{trust}$\\ \hline\hline
IC\,1515	&	11.1	&	14.5	&	16.7	&	25.6	&	30.0	&	44.5	&	44.5		\\
IC\,1516	&	24.8	&	43.1	&	44.2	&	53.9	&		&		&	99.3		\\
NGC\,450	&	3.9	&	9.0	&	9.7	&	14.2	&	15.5	&	18.1	&	18.3		\\
NGC\,493	&	6.3	&	12.6	&	13.9	&	22.7	&	27.1	&	41.7	&	42.5		\\
NGC\,497	&	12.6	&	20.1	&	26.4	&	50.3	&	67.9	&	88.0	&	90.6		\\
NGC\,799	&	8.1	&	17.1	&	21.3	&	31.6	&	37.6	&	47.0	&	48.7		\\
NGC\,856	&	10.5	&	41.9	&		&		&		&		&	41.9		\\
NGC\,941	&	4.0	&	7.3	&	8.9	&	14.5	&		&		&	29.0		\\
NGC\,1090	&	21.2	&	26.8	&	29.0	&	39.1	&		&		&	46.0		\\
NGC\,7398	&	6.0	&	8.0	&	10.0	&	17.9	&	21.9	&	39.8	&	39.8		\\
UGC\,139	&	10.7	&	16.6	&	19.5	&	24.4	&	26.3	&	45.8	&	45.8		\\
UGC\,272	&	2.7	&	6.4	&	10.1	&	15.6	&	18.3	&	30.2	&	30.2		\\
UGC\,651	&	3.2	&	8.6	&	9.7	&	15.0	&	16.1	&	25.8	&	25.8		\\
UGC\,866	&	2.8	&	6.9	&	7.9	&	9.8	&		&		&	13.1		\\
UGC\,1934	&	20.7	&	31.0	&	31.0	&	33.0	&	54.7	&	86.7	&	91.8		\\
UGC\,2081	&	5.0	&	11.6	&	13.2	&	19.0	&		&		&	19.0		\\
UGC\,2311	&	10.4	&	14.5	&	17.6	&	27.0	&	29.0	&	33.2	&	46.7		\\
UGC\,2319	&	2.9	&	8.8	&	10.8	&	22.5	&	24.5	&	32.3	&	40.2		\\
UGC\,2418	&	2.9	&	8.8	&	11.8	&	27.4	&	28.4	&	31.9	&	44.1		\\
UGC\,12183	&	1.2	&	4.6	&	5.8	&	13.8	&	15.0	&	28.8	&	32.6		\\
UGC\,12208	&	2.4	&	5.9	&	7.1	&	11.9	&	14.3	&	16.6	&	18.5		\\
UGC\,12709	&	4.8	&	10.9	&	13.3	&	18.2	&	19.4	&	24.2	&	25.7		\\ \hline \hline

\end{tabular}
\caption[Fit regions]{Fit regions in [kpc]}\label{tbl:fitregions}
\end{table}

\subsection{Rectified Polar Plots}\label{sec:PolarPlots}
One possible reason that truncations in face-on galaxies are 
in principle difficult to see is 
that galaxies do not remain circularly symmetric, but become ragged or 
lopsided towards the faint outskirts. Any method of analysis that 
involves azimuthal averaging or assumptions on azimuthal symmetry would
smooth such features out, since no galaxy is perfectly symmetric 
\citet{kf11}.  \citet{pdla02} plotted 
the isophotes of the galaxy {NGC\,5923} in polar coordinates (their Figure 9),
which illustrates this. These authors demonstrated that taking individual
sliced rows in this polar plot did not significantly impact the inner slope
or blur the outer slope. We feel however that at the very faint outskirts
of galaxies as we are studying this effect may still be significant.

Here, we will continue and expand upon the work by \citet{pdla02}.
First, we project all galaxies into polar coordinates.
We note that the outskirts of most galaxies indeed suffer from 
irregularities, which would smooth out possible truncations in any 
ellipse-averaging scheme. 

We therefore attempt to `rectify' the polar plots (RPP), such that the 
outer radii become well behaved.
A boxcar smoothing was applied to the data, such that in the 
azimuthal direction the data are smoothed by a window of length 
$15^\circ$, one centered on one side of the major axis and then moving 
arounds the galaxy. In the radial direction the smoothing is effectively
the same as the pixel size, so no radial details would be lost.
We define various target brightness levels for rectification, 
specifically we tried every half magnitude between 25 and 27 
$r'$-mag/arcsec$^2$. 
Along each radial sector-profile, 
we measure the first crossing of this rectification target brightness.
The median over all azimuths of this first crossing radius 
yields a target rectification radius.
Each sector of fixed azimuth in the original image is then 
stretched or compressed, such that the radius of the first 
crossing as seen in the smoothed image matches the target rectification radius.

The polar plot is partitioned into 12 azimuthal sectors in order 
to deal with problems in the image.
In cases where the sector suffered from issues, for example due to 
masking, the entire sector was excluded from the analysis. The procedure
has been illustrated for two galaxies, IC\,1515 and UGC\,12208 in Figure 
\ref{fig:PolarPlots}. The dark levels at larger radii shows that the noise  
illustrates the goodness of the fit for the background; the white 
areas result from the borders of the image and from bright
stars that have been masked in the sky plot. 
The rectification has been performed here at 26 mag/arcsec$^{2}$, which 
corresponds to the yellow and light-green in the figure.
Comparing the left- and right-hand panels (focus for example on the structure and curvature of the
transition from the yellow to the orange coded levels)  clearly 
shows the effects of the rectification at fainter levels. IC\,1515 shows some 
lopsidedness at faint levels (larger extent in angles $\pm \pi$ at the left)
that has been rectified in the process.

The final combined profiles are shown Figure \ref{fig:Combined} for the
same two galaxies. Obviously only sectors that are not affected by 
difficuties such as background level or masked stars have been included.
It shows PAS in red, the conventional ellipse fitting
in blue,  EP in green and RPP in purple, the latter for rectification target 
brightness 
levels of 25, 26 and 27 $r'$-mag/arcsec$^2$. Some offsets are applied as 
explained in the caption. The dashed lines have been explained in the caption.
We present a full set of illustrations for the full set of galaxies in our sample 
in the online appendix. In that appendix we show the RPP results only for 
25, 26 and 27 $r'$-mag/arcsec$^2$.

\begin{table*}
\centering
\begin{tabular}{l|cc|cc|cc|c}
 & \multicolumn{2}{ c|}{Break}& \multicolumn{2}{ c| }{Truncation}& \multicolumn{2}{c  }{Halo}\\
Galaxy & $R$ & $\mu$& $R$ & $\mu$& $R$ & $\mu$&$R_{25}$\\ \hline\hline
IC\,1515	&$	16.5	^{+	0.4	}_{-	0.4	}$&$	23.6	^{+	0.1	}_{-	0.1	}$	&								&								&	$	32.1	^{+	1.0	}_{-	1.3	}$&$	29.0	^{+	0.2	}_{-	0.3	}$	&		21.3	\\
IC\,1516	&						&							&	$	45.4	^{+	2.7	}_{-	4.9	}$	&	$	28.0	^{+	0.5	}_{-	0.8	}$	&							&							&		27.5	\\
NGC\,450	&$	9.6	^{+	0.1	}_{-	0.1	}$&$	23.9	^{+	0.0	}_{-	0.0	}$	&								&								&	$	14.5	^{+	0.5	}_{-	0.6	}$&$	28.1	^{+	0.3	}_{-	0.4	}$	&		11.1	\\
NGC\,493	&$	13.0	^{+	0.3	}_{-	0.4	}$&$	23.0	^{+	0.1	}_{-	0.1	}$	&								&								&	$	25.3	^{+	0.9	}_{-	0.9	}$&$	28.4	^{+	0.2	}_{-	0.2	}$	&		17.9	\\
NGC\,497	&$	22.8	^{+	1.2	}_{-	1.3	}$&$	22.3	^{+	0.1	}_{-	0.1	}$	&								&								&	$	61.2	^{+	1.9	}_{-	2.0	}$&$	28.7	^{+	0.2	}_{-	0.2	}$	&		40.9	\\
NGC\,799	&$	20.3	^{+	0.4	}_{-	0.5	}$&$	23.5	^{+	0.1	}_{-	0.1	}$	&								&								&	$	37.5	^{+	1.7	}_{-	3.1	}$&$	29.4	^{+	0.4	}_{-	0.7	}$	&		24.9	\\
NGC\,856	&						&							&								&								&							&							&		21.1	\\
NGC\,941	&$	7.3	^{+	0.2	}_{-	0.2	}$&$	24.1	^{+	0.1	}_{-	0.1	}$	&								&								&							&							&		8.9	\\
NGC\,1090	&$	26.1	^{+	0.2	}_{-	0.3	}$&$	26.1	^{+	0.2	}_{-	0.3	}$	&								&								&							&							&		21.9	\\
NGC\,7398	&$	8.5	^{+	0.3	}_{-	0.4	}$&$	22.8	^{+	0.0	}_{-	0.0	}$	&								&								&	$	16.2	^{+	2.2	}_{-	1.6	}$&$	26.1	^{+	0.6	}_{-	0.5	}$	&		15.3	\\
UGC\,139	&$	18.3	^{+	0.9	}_{-	1.4	}$&$	24.9	^{+	0.2	}_{-	0.2	}$	&								&								&	$	29.2	^{+	2.1	}_{-	2.6	}$&$	29.4	^{+	0.4	}_{-	0.7	}$	&		19.1	\\
UGC\,272	&$	7.9	^{+	0.7	}_{-	1.0	}$&$	22.9	^{+	0.2	}_{-	0.2	}$	&								&								&	$	16.6	^{+	6.8	}_{-	4.8	}$&$	27.2	^{+	2.2	}_{-	1.9	}$	&		13.3	\\
UGC\,651	&$	9.0	^{+	0.4	}_{-	0.6	}$&$	23.9	^{+	0.1	}_{-	0.2	}$	&								&								&	$	13.6	^{+	3.4	}_{-	1.9	}$&$	27.0	^{+	1.4	}_{-	0.9	}$	&		11.6	\\
UGC\,866	&						&							&								&								&	$	9.3	^{+	0.2	}_{-	0.2	}$&$	29.0	^{+	0.1	}_{-	0.1	}$	&		5.1	\\
UGC\,1934	&$	25.8	^{+	1.7	}_{-	2.5	}$&$	23.5	^{+	0.3	}_{-	0.4	}$	&								&								&	$	40.9	^{+	1.3	}_{-	1.2	}$&$	26.6	^{+	0.1	}_{-	0.1	}$	&		35.1	\\
UGC\,2081	&$	12.3	^{+	0.3	}_{-	0.3	}$&$	25.6	^{+	0.1	}_{-	0.1	}$	&								&								&							&							&		10.6	\\
UGC\,2311	&$	17.4	^{+	0.3	}_{-	0.5	}$&$	23.8	^{+	0.1	}_{-	0.1	}$	&								&								&	$	32.8	^{+	1.5	}_{-	1.8	}$&$	28.8	^{+	0.3	}_{-	0.4	}$	&		21.7	\\
UGC\,2319	&$	10.8	^{+	0.1	}_{-	0.2	}$&$	22.4	^{+	0.0	}_{-	0.0	}$	&								&								&	$	27.1	^{+	0.6	}_{-	0.9	}$&$	28.1	^{+	0.1	}_{-	0.2	}$	&		19.0	\\
UGC\,2418	&$	9.6	^{+	0.2	}_{-	0.2	}$&$	22.1	^{+	0.0	}_{-	0.0	}$	&								&								&	$	25.4	^{+	1.7	}_{-	2.4	}$&$	28.1	^{+	0.4	}_{-	0.7	}$	&		17.7	\\
UGC\,12183	&$	5.3	^{+	0.4	}_{-	0.5	}$&$	22.8	^{+	0.1	}_{-	0.2	}$	&								&								&	$	17.4	^{+	1.2	}_{-	1.2	}$&$	28.9	^{+	0.4	}_{-	0.4	}$	&		10.0	\\
UGC\,12208	&$	5.6	^{+	0.1	}_{-	0.1	}$&$	22.7	^{+	0.0	}_{-	0.0	}$	&	$	13.4	^{+	0.9	}_{-	2.9	}$	&	$	27.7	^{+	0.6	}_{-	1.9	}$	&							&							&		9.3	\\
UGC\,12709	&$	12.7	^{+	0.1	}_{-	0.1	}$&$	24.8	^{+	0.0	}_{-	0.0	}$	&	$	19.6	^{+	0.7	}_{-	1.3	}$	&	$	27.9	^{+	0.3	}_{-	0.6	}$	&							&							&		13.0	\\
\hline\hline
\end{tabular}
\caption[Position and brightness of the various features]{The position $R$ and surface brightness $\mu$ at the onset of the haloes, breaks and truncations in our galaxy sample. The positions are in kpc. The surface brightness is in $r'$-mag/
arcsec$^2$. }\label{tbl:main}
\end{table*}

\begin{table*}
\centering
\begin{tabular}{l|cc|cc|cc|cc}
 & \multicolumn{2}{ c|}{Inner galaxy}& \multicolumn{2}{ c| }{Outer galaxy}& \multicolumn{2}{c|  }{Truncation}& \multicolumn{2}{c  }{Halo}\\ 
 Galaxy & $h$ & $\mu_0$& $h$ & $\mu_0$& $h$ & $\mu_0$& $h$ & $\mu_0$\\\hline\hline
IC\,1515	&$	5.8	^{+	0.1	}_{-	0.1	}$&$	20.5	^{+	0.0	}_{-	0.0	}$&$	3.1	^{+	0.1	}_{-	0.1	}$&$	17.9	^{+	0.2	}_{-	0.3	}$	&						&							&$	16.9	^{+	5.4	}_{-	4.1	}$&$	26.9	^{+	0.6	}_{-	0.8	}$		\\
IC\,1516	&						&						&$	6.4	^{+	0.2	}_{-	0.2	}$&$	20.3	^{+	0.2	}_{-	0.2	}$	&$	4.3	^{+	1.0	}_{-	0.7	}$&$	16.6	^{+	2.4	}_{-	2.6	}$	&														\\
NGC\,450	&$	3.7	^{+	0.1	}_{-	0.1	}$&$	21.2	^{+	0.0	}_{-	0.0	}$&$	1.3	^{+	0.0	}_{-	0.1	}$&$	15.8	^{+	0.3	}_{-	0.4	}$	&						&							&$	5.2	^{+	1.5	}_{-	1.1	}$&$	25.1	^{+	0.9	}_{-	1.0	}$		\\
NGC\,493	&$	6.0	^{+	0.2	}_{-	0.1	}$&$	20.6	^{+	0.0	}_{-	0.0	}$&$	2.5	^{+	0.1	}_{-	0.1	}$&$	17.3	^{+	0.3	}_{-	0.3	}$	&						&							&$	13.6	^{+	2.9	}_{-	1.9	}$&$	26.4	^{+	0.5	}_{-	0.5	}$		\\
NGC\,497	&$	13.1	^{+	1.1	}_{-	0.9	}$&$	20.5	^{+	0.1	}_{-	0.1	}$&$	6.6	^{+	0.2	}_{-	0.2	}$&$	18.6	^{+	0.2	}_{-	0.2	}$	&						&							&$	28.5	^{+	2.1	}_{-	3.5	}$&$	26.3	^{+	0.2	}_{-	0.4	}$		\\
NGC\,799	&$	17.3	^{+	1.2	}_{-	1.0	}$&$	22.3	^{+	0.1	}_{-	0.0	}$&$	3.2	^{+	0.2	}_{-	0.2	}$&$	16.5	^{+	0.5	}_{-	0.7	}$	&						&							&$	14.0	^{+	4.9	}_{-	5.2	}$&$	26.6	^{+	0.9	}_{-	2.1	}$		\\
NGC\,856	&						&						&$	5.2	^{+	0.0	}_{-	0.0	}$&$	20.7	^{+	0.0	}_{-	0.0	}$	&						&																					\\
NGC\,941	&$	2.5	^{+	0.1	}_{-	0.1	}$&$	20.8	^{+	0.1	}_{-	0.1	}$&$	1.6	^{+	0.0	}_{-	0.0	}$&$	19.0	^{+	0.1	}_{-	0.1	}$	&						&							&														\\
NGC\,1090	&$	6.1	^{+	0.1	}_{-	0.1	}$&$	21.1	^{+	0.1	}_{-	0.1	}$&$	3.6	^{+	0.3	}_{-	0.3	}$&$	17.7	^{+	0.9	}_{-	0.8	}$	&						&							&														\\
NGC\,7398	&$	14.3	^{+	1.0	}_{-	0.8	}$&$	22.1	^{+	0.0	}_{-	0.0	}$&$	2.5	^{+	0.3	}_{-	0.3	}$&$	19.1	^{+	0.5	}_{-	0.5	}$	&						&							&$	6.7	^{+	0.7	}_{-	0.5	}$&$	23.5	^{+	0.5	}_{-	0.4	}$		\\
UGC\,139	&$	6.1	^{+	0.2	}_{-	0.2	}$&$	21.6	^{+	0.1	}_{-	0.1	}$&$	2.6	^{+	0.4	}_{-	0.4	}$&$	17.3	^{+	1.2	}_{-	1.6	}$	&						&							&$	14.7	^{+	6.0	}_{-	4.5	}$&$	27.2	^{+	0.9	}_{-	1.3	}$		\\
UGC\,272	&$	4.7	^{+	0.0	}_{-	0.0	}$&$	21.1	^{+	0.0	}_{-	0.0	}$&$	2.2	^{+	0.4	}_{-	0.5	}$&$	19.1	^{+	0.7	}_{-	1.3	}$	&						&							&$	4.6	^{+	2.0	}_{-	0.7	}$&$	23.2	^{+	2.3	}_{-	1.3	}$		\\
UGC\,651	&$	3.3	^{+	0.0	}_{-	0.0	}$&$	20.9	^{+	0.0	}_{-	0.0	}$&$	1.6	^{+	0.4	}_{-	0.3	}$&$	17.9	^{+	1.2	}_{-	1.6	}$	&						&							&$	4.3	^{+	1.5	}_{-	0.6	}$&$	23.6	^{+	1.6	}_{-	0.9	}$		\\
UGC\,866	&						&						&$	1.1	^{+	0.0	}_{-	0.0	}$&$	20.2	^{+	0.1	}_{-	0.1	}$	&						&							&$	16.0	^{+	5.0	}_{-	7.1	}$&$	28.4	^{+	0.2	}_{-	0.5	}$		\\
UGC\,1934	&$	6.8	^{+	0.0	}_{-	0.0	}$&$	19.4	^{+	0.0	}_{-	0.0	}$&$	5.2	^{+	0.3	}_{-	0.3	}$&$	18.1	^{+	0.4	}_{-	0.4	}$	&						&							&$	22.8	^{+	1.4	}_{-	1.2	}$&$	24.7	^{+	0.2	}_{-	0.2	}$		\\
UGC\,2081	&$	3.3	^{+	0.0	}_{-	0.0	}$&$	21.5	^{+	0.0	}_{-	0.0	}$&$	1.7	^{+	0.1	}_{-	0.1	}$&$	17.6	^{+	0.4	}_{-	0.4	}$	&						&							&														\\
UGC\,2311	&$	7.3	^{+	0.4	}_{-	0.3	}$&$	21.2	^{+	0.1	}_{-	0.1	}$&$	3.3	^{+	0.1	}_{-	0.1	}$&$	18.0	^{+	0.2	}_{-	0.2	}$	&						&							&$	17.5	^{+	6.0	}_{-	9.3	}$&$	26.8	^{+	0.6	}_{-	2.4	}$		\\
UGC\,2319	&$	5.3	^{+	0.0	}_{-	0.0	}$&$	20.2	^{+	0.0	}_{-	0.0	}$&$	3.1	^{+	0.1	}_{-	0.1	}$&$	18.6	^{+	0.1	}_{-	0.1	}$	&						&							&$	13.2	^{+	3.3	}_{-	2.7	}$&$	25.9	^{+	0.5	}_{-	0.7	}$		\\
UGC\,2418	&$	7.7	^{+	0.3	}_{-	0.3	}$&$	20.8	^{+	0.0	}_{-	0.0	}$&$	2.9	^{+	0.1	}_{-	0.1	}$&$	18.5	^{+	0.2	}_{-	0.2	}$	&						&							&$	8.0	^{+	2.7	}_{-	1.7	}$&$	24.7	^{+	1.1	}_{-	1.2	}$		\\
UGC\,12183	&$	2.9	^{+	0.0	}_{-	0.0	}$&$	20.8	^{+	0.0	}_{-	0.0	}$&$	2.1	^{+	0.1	}_{-	0.1	}$&$	20.1	^{+	0.1	}_{-	0.1	}$	&						&							&$	6.3	^{+	1.2	}_{-	0.8	}$&$	25.9	^{+	0.7	}_{-	0.7	}$		\\
UGC\,12208	&$	6.6	^{+	0.4	}_{-	0.3	}$&$	21.7	^{+	0.0	}_{-	0.0	}$&$	1.7	^{+	0.0	}_{-	0.0	}$&$	19.0	^{+	0.1	}_{-	0.1	}$	&$	0.8	^{+	0.4	}_{-	0.2	}$&$	19.0	^{+	0.1	}_{-	0.1	}$	&														\\
UGC\,12709	&$	7.9	^{+	0.2	}_{-	0.2	}$&$	23.0	^{+	0.0	}_{-	0.0	}$&$	2.4	^{+	0.1	}_{-	0.1	}$&$	19.0	^{+	0.2	}_{-	0.2	}$	&$	1.5	^{+	0.2	}_{-	0.2	}$&$	19.0	^{+	0.2	}_{-	0.2	}$	&														\\
\hline\hline
\end{tabular}
\caption[Fitted scale lengths]{Fitted scale lengths in kpc and central brightness in $r'$-mag/arcsec$^2$ as measured from the ellipse profiles}\label{tbl:scalelengths}
\end{table*}

\section{Results}\label{sec:results}
\subsection{Measurement Strategy}
Each set of profiles was inspected for the presence of breaks, 
truncations and haloes. 
We classified a feature as a break when there is a significant 
change in scale length inside and outside the radius of the feature. 
In almost every case the break can be associated directly with a 
feature in the galaxy, such as a bulge, bar or spiral arm.
Breaks often occur in the inner, brighter parts of the galaxy, 
well above the pixel-to-pixel noise, and as such, they are visible 
in all four types of profiles.
The rectified polar profiles are intended for detecting truncations.
We classified truncations of the profiles, when at levels fainter
than 26 mag arcsec$^{-2}$
the profile definitely changes slope and drops towards the sky
before disappearing in the noise. NGC\,1090 has a break at 26.1 mag 
arcsec$^{-2}$, but has a very minor change in slope. We do not classify it as 
a truncation, but a break instead.
Haloes are the opposite of truncations; when rather than steepening, 
the profile becomes less steep. 
Haloes should be visible in both the rectified polar profiles and in 
the ellipse profiles.
As we expect truncations and haloes to occur at the outer edges of 
the profiles, deep below the pixel-to-pixel noise, the PAS and EP 
methods are not used.

To ensure unbiased results from the data, our team split up into groups.
Each group went through all of the profiles and classified what they 
believed to be breaks, truncations and/or haloes. 
After that, we joined up again to discuss our findings and agree on 
any difficult cases.

We then set out to quantify our results by measuring the parameters 
of each feature. 
We defined fit-regions to the profile on both sides of a feature 
(shown in Table \ref{tbl:fitregions}).
In each of  these regions we have a set of radii $R$ and observed 
surface brightness levels $\mu_\textrm{obs}(R)$, to which we fit a 
linear relation with scale length $h$ and central brightness $\mu_0$ 
at $R=0$ of form
\begin{equation}
\mu(R;\mu_0,h) = \mu_0 + 1.086 \frac{R}{h}\,\,.
\end{equation}
The fits were performed on the ellipse-fitting profiles, as
the EP profiles in general do 
not extend sufficiently far out, the PAS profiles suffer from convolution and 
the RPP results have been produced by stretching and compressing the data.

We give the fitting algorithm the freedom to estimate the noise 
by itself, in a method inspired by the use of the jitter parameter in 
\citet{Hou2012A}. 
We assume that the probability density $p$ for parameters $\mu_0$ and $h$ 
is given by the Gaussian likelihood function
\begin{equation}
p =\prod_{R_i}^{R_N} \frac{1}{\sqrt{2\pi\sigma^2}}\exp{\left(\frac{-[\mu(R_i) - \mu_\textrm{obs}(R_i)]^2}{2\sigma^2}\right)}\,\,,
\end{equation}
where $R_N$ is the collection of all measurements at positions $R_i$ 
in the fitting region. It was assumed that the dispersion $\sigma $ 
is not a function of radius along the profiles.
The log-likelihood $\ln p$ is then 
\begin{eqnarray}
\ln p &=& -\frac{1}{2}\sum_{R_i}^{R_N}\left( \frac{[\mu(R_i;\mu_0,h) - \mu_\textrm{obs}(R_i)]^2}{2\sigma^2}n\right.\nonumber\\
&& \left. + \ln{\left(2\pi\sigma^2\right)}\right)\,\,,\\
            &\approx& -\frac{1}{2}\sum_{R_i}^{R_N}\left(\frac{[\mu(R_i;\mu_0,h) - \mu_\textrm{obs}(R_i)]^2}{2\sigma^2}\right.\nonumber\\
&& \left. + \ln{(\sigma^2)}\right)\,\,,
\end{eqnarray}
where we drop the $2\pi$ term as it is only a constant and does not 
influence the outcome. Setting $\ln p = -1/2 \chi^2$, we then have 
\begin{equation}
\chi^2 = \sum_{R_i}^{R_N}\left( \frac{[\mu(R_i;\mu_0,h) - \mu_\textrm{obs}(R_i)]^2}{2\sigma^2} + \ln{(\sigma^2)}\right)\,\,.
\end{equation}

\begin{figure*}
\centering
\includegraphics[width=0.48\textwidth]{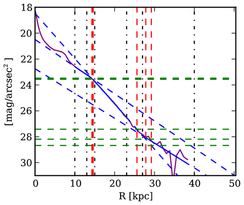}
\includegraphics[width=0.48\textwidth]{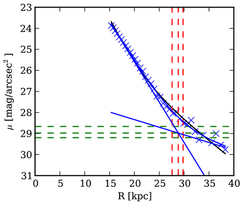}
\caption[Scale length measurement strategy]{Left panel shows the fitting of the profile of IC\,1515 in three different sections, for a break and a halo. The red vertical dashed lines denote the estimation of the position $R$ and uncertainty for each feature. The green horizontal lines show the central position and errors on the brightness $\mu$. The vertical black dash-dotted lines show the outer boundaries of the fit regions. The right panel shows the improved fit to the halo. See the text for more details.}
\label{fig:fits}
\end{figure*}

We perform an initial fit to the data by minimizing $\chi^2$, using the 
\textsc{Python} routine \textsc{scipy.optimize.leastsq}. 
After we had achieved a proper fit, we switched to a 
Markov-Chain Monte-Carlo (MCMC) sampler, \textsc{emcee}, to  estimate 
the posterior likelihood distribution $\ln p$ and the distribution in 
each of the parameters \citep{emcee}.
We used 300 separate walkers, which after a burn-in period of 50 
iterations, sample the data for 500 iterations, giving  135000 samples.
For each parameter, the median value ($50\%$) of that parameter 
distribution is used as the reported value, while the upper and 
lower errors are reported based on the $16\%$ and $84\%$ parts 
of the distribution.
The radii of the breaks and truncations were calculated using 
the intersection of both sets of samples, with the error derived 
in the same way as before. These radii were used
to calculate the brightness, which 
is the brightness of the profile at the onset of that feature at position $R$.
We demonstrate this in the left panel of Figure \ref{fig:fits}.

This method was insufficient for the halo, as the halo lies `on top' 
of the regular stellar disc profile. 
We thus fit the combined profile of both the stellar profile ($\mu_\textrm{0,s}, h_\textrm{s}$) and the halo ($\mu_\textrm{0,h}, h_\textrm{h}$). 
For this, we used the combination of both fitting regions, as well as 
the region between these two.
The combined profile is given by 
\begin{equation}
 \mu(R; \mu_\textrm{0,s}, h_s, \mu_\textrm{0,h}, h_\textrm{h}) = \ln \left[ e^{\mu(R;\mu_\textrm{0,s},h_\textrm{s})} + e^{\mu(R;\mu_\textrm{0,h},h_\textrm{h})} \right].
\end{equation}
We used the parameters from the individual line segment fits as initial 
values and again performed the subsequent MCMC analysis using \textsc{emcee}.
The brightness $\mu$ of the halo onset was calculated from the brightness 
of only the halo, rather than from the combination of halo and stellar disc.
We demonstrate the new combined fit in the right panel of Figure \ref{fig:fits}.
The combined fits for all profiles can be found in the Appendix in Figures 
A23 and A24.

The total luminosity of the galaxy $L_\textrm{total}$ was measured by 
integrating the profile $\mu_0(R)$ (in counts) up to the trust radius 
$R_\textrm{trust}$:
\begin{equation}
L_\textrm{total} = \int_{R=0}^{R_\textrm{trust}} 2 \pi R (1-e) 10^{-0.4\mu_\textrm{obs}(R)} \,\textrm{d}R\,\,.
\end{equation}
The ellipticity term  $e$ is used to correct for the inclination.
The total luminosity ($R\rightarrow\infty$) of the flattened halo 
$L_\textrm{halo}$ is 
\begin{equation}
L_\textrm{halo} = 2 \pi h_\textrm{h}^2 (1-e) 10^{-0.4\mu_\textrm{0,h}}\,\,.\label{eqn:haloinfty}
\end{equation}
Here $h$ and $\mu_0$ are the scale length and central brightness 
as measured for the halo.
We are interested in what fraction $\eta$ of the total luminosity 
originates from the halo. 
Because $L_\textrm{halo}$ as defined in Equation \ref{eqn:haloinfty} 
runs to infinity, while $L_\textrm{total}$ only integrates up to 
$R_\textrm{trust}$, we need to correct for this missing light, as follows:
\begin{eqnarray}
\eta &=& \frac{L_\textrm{halo}}{L_\textrm{total}} \,\,\,[R\rightarrow \infty], \\
     &=& \frac{L_\textrm{halo}}{L_\textrm{total}\,[R\leq R_\textrm{trust}] + L_\textrm{halo}\,[R>R_\textrm{trust}]} \,\,\,[R\rightarrow \infty],\nonumber
\end{eqnarray}
where we assume that the only light beyond $R_\textrm{trust}$ originates from 
the halo component, such that
\begin{eqnarray}
L_\textrm{halo} &=& 2\pi h_\textrm{h} 10^{-0.4\mu_\textrm{0,h}} \exp\left(-\frac{R_\textrm{trust}}{h_\textrm{h}}\right) (h_\textrm{h} + R_\textrm{trust})\nonumber\\
&&[R>R_\textrm{trust}]\,\,.
\end{eqnarray}
The radii and brightness levels for the breaks, haloes and truncations 
are shown in Table \ref{tbl:main}, the measured scale lengths are shown in 
Table \ref{tbl:scalelengths} and the halo light fractions $\eta$ are shown 
in Table \ref{tbl:eta}.

\begin{table}
\centering
\begin{tabular}{l|cc|c}
Galaxy & Morphological type & $M_{\rm abs}$ & $\eta$\\\hline \hline
IC\,1515	&	Sb	&	-21.9	&$	0.020	^{+	0.005	}_{-	0.007	}$	\\
IC\,1516	&	Sbc	&	-22.2								\\
NGC\,450	&	SABc	&	-19.8	&$	0.058	^{+	0.020	}_{-	0.022	}$	\\
NGC\,493	&	Sc	&	-20.4	&$	0.031	^{+	0.006	}_{-	0.006	}$	\\
NGC\,497	&	SBbc	&	-22.7	&$	0.025	^{+	0.002	}_{-	0.004	}$	\\
NGC\,799	&	Sba	&	-22.0	&$	0.019	^{+	0.007	}_{-	0.016	}$	\\
NGC\,856	&	SABc	&	-21.8								\\
NGC\,941	&	SABc	&	-19.5								\\
NGC\,1090	&	Sbc	&	-21.3								\\
NGC\,7398	&	SABc	&	-21.0	&$	0.152	^{+	0.031	}_{-	0.023	}$	\\
UGC\,139	&	SABc	&	-20.5	&$	0.023	^{+	0.008	}_{-	0.012	}$	\\
UGC\,272	&	SABc	&	-19.6	&$	0.176	^{+	0.161	}_{-	0.088	}$	\\
UGC\,651	&	Sc	&	-19.2	&$	0.155	^{+	0.096	}_{-	0.055	}$	\\
UGC\,866	&	Sd	&	-17.2	&$	0.123	^{+	0.009	}_{-	0.021	}$	\\
UGC\,1934	&	Sbc	&	-21.9	&$	0.094	^{+	0.007	}_{-	0.007	}$	\\
UGC\,2081	&	SABc	&	-19.0								\\
UGC\,2311	&	Sb	&	-22.0	&$	0.023	^{+	0.005	}_{-	0.023	}$	\\
UGC\,2319	&	Sc	&	-20.7	&$	0.038	^{+	0.008	}_{-	0.010	}$	\\
UGC\,2418	&	SABb	&	-21.3	&$	0.056	^{+	0.025	}_{-	0.026	}$	\\
UGC\,12183	&	Sbc	&	-19.5	&$	0.049	^{+	0.014	}_{-	0.013	}$	\\
UGC\,12208	&	Scd	&	-19.7								\\
UGC\,12709	&	SABm	&	-19.1								\\
\hline\hline
\end{tabular}
\caption[Absolute brightness for each galaxy and the halo light fraction]{Absolute brightness for each galaxy and the halo light fraction $\eta$. The values for $M_{\rm abs}$ refetr to the $r'$-band.}\label{tbl:eta}
\end{table}

\subsection{Notes on Individual Galaxies}
The vast majority (91\%) of our sample shows some form of break in their 
light profile. 
These can be ``classical'' breaks ($86\%$), and/or truncations ($14\%$) 
at much lower surface brightness levels, with only NGC~856 and UGC~866 
showing no indication of a break or a  truncation.
The most distinct truncation in the sample is seen in UGC~12208.
Only NGC\,856 has a pure exponential disc.

IC\,1515, IC\,1516, and NGC\,799 have the lowest inclinations of the sample. 
Between the three of them, breaks, haloes, and truncations have all been 
detected, explicitly showing the feasibility of detecting these features 
in face-on galaxies.
Conversely, all the highly inclined ($i >$ 70 deg) galaxies of our sample 
(NGC\,493, UGC\,651, UGC\,866, UGC\,1934, and UGC\,2319),  feature 
apparent stellar haloes yet none displays a truncation.

Four galaxies from our sample overlap with the sample of \citet{pt06}.
For NGC\,450, NGC\,941, and UGC\,12709 both studies agree on the occurrence 
of a break as well as the radius at which it is observed. 
For UGC2081, there is a disagreement; we did not detect the break they find 
at  $53''$. However, they did note that ``downbending break'' 
starts ``clearly at ~70 arcsec''.  
This galaxy was also analyzed by \citet{bt13}, who report a break at 
$66.5''$ from $r'$-band profiles, while we report $71''$. 

Further agreement with the literature is found for the barred spiral 
UGC\,2311. The break found at $37''$ is broadly consistent with 
\citet{bt13}, who report it at $42''$.
Additionally, NGC\,941 was also analyzed by \citet{mm13} in the 
mid-infrared at 3.6 $\mu$, where they report a break at $80''$, 
while we report $61''$.

\section{Discussion}\label{sec:discussion}

\subsection{Comparing the Various Techniques}
We have used four different methods to analyze the profiles. 
Ellipse profiles have a well-established history and set the standard for 
surface photometry. The EP, which was already used long ago, and 
PAS methods were presented, applied and discussed in \citet{pkj13}. 
The EP is found to reproduce similar results as ellipse fitting in 
bright regions of the galaxy, but suffers more from background noise.
The PAS method was developed as a way to project face-on galaxies 
to edge-on in order to detect truncations.
The PAS method was found by \cite{pkj13} to create a differently shaped profile.
However, using these new profiles \cite{pkj13} were unable to reproduce 
the correlations seen in the edge-on sample of \citet{mbt12}, and so argued 
that no real truncations had been detected in the \cite{pkj13} sample. 
In this paper, we introduce a fourth method, rectified polar 
profiles, inspired by the earlier work by \citet{pdla02}.

The images analyzed in this paper are two magnitudes deeper than the 
default SDSS images.
The effect of this is clearly noticeable on the EP and PAS methods. 
These now extend to further radii and fainter surface brightness levels, 
although the pixel-to-pixels noise  remains the limiting factor.
As previously seen in \citet{pkj13}, the ellipse method remains superior 
at faint brightness levels, as the averaging over large areas allows it 
to measure reliably the profile below the pixel-to-pixel noise level.
Similar to this, the polar rectified profiles also average over large areas 
and can thus reach the same limiting brightness levels.

Comparing the polar and ellipse profiles directly, we note that the results 
are often similar.
This is expected, as non-rectified polar profiles are effectively the same 
as ellipse profiles.
At  faint brightness levels, the outer edges of galaxies are clearly wobbling, 
as previously noted by \citet{pdla02}.
It was postulated by \citet{kf11} that these local variations are important 
when hunting for truncations.
Our work here demonstrates this, as can be observed in Figure A1 (Appendix), 
where the ellipse profile displays a bump in the profile, yet the rectified 
polar plots show that the profile is actually dropping towards the sky at a 
constant rate.

In conclusion, we find that the PAS and EP methods are ill suited for the 
detection of truncations. They should only be used in brighter parts of 
galaxies. Rectified polar and ellipse profiles are on par in terms of the 
limiting magnitude. The rectified polar profiles demonstrate that it is 
important to take the ragged or lopsided nature of galaxies into account, 
as the ellipse method is averaging these features out.

\subsection{Possible PSF Effects}\label{sec:psfdiscussion}

\begin{figure*}
\centering
\includegraphics[width=0.32\textwidth]{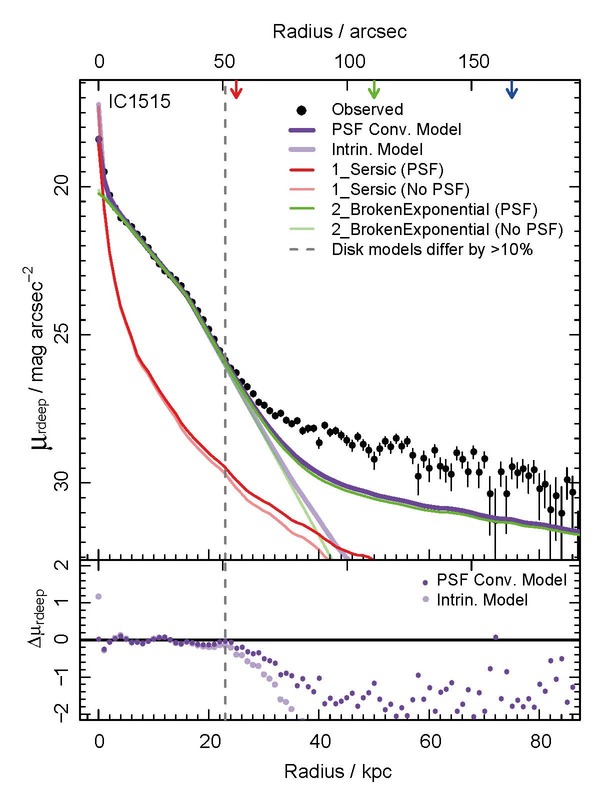}
\includegraphics[width=0.32\textwidth]{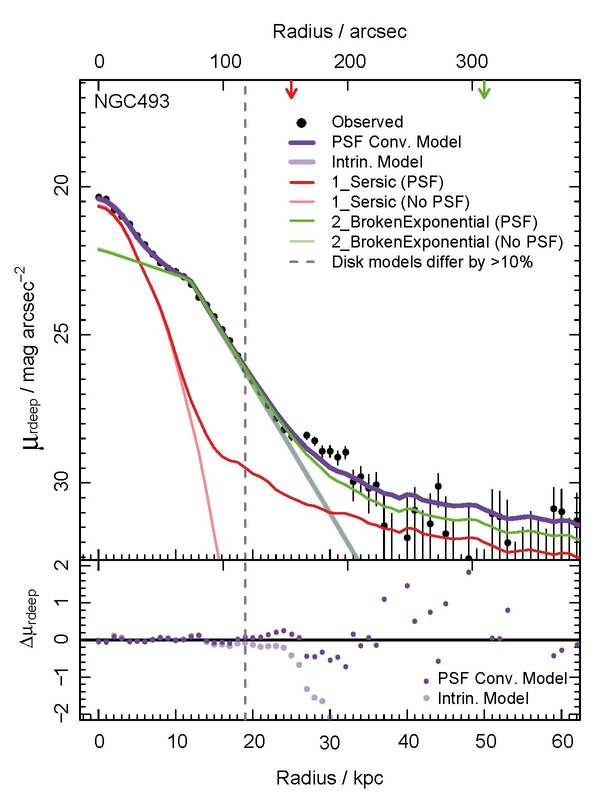}
\includegraphics[width=0.32\textwidth]{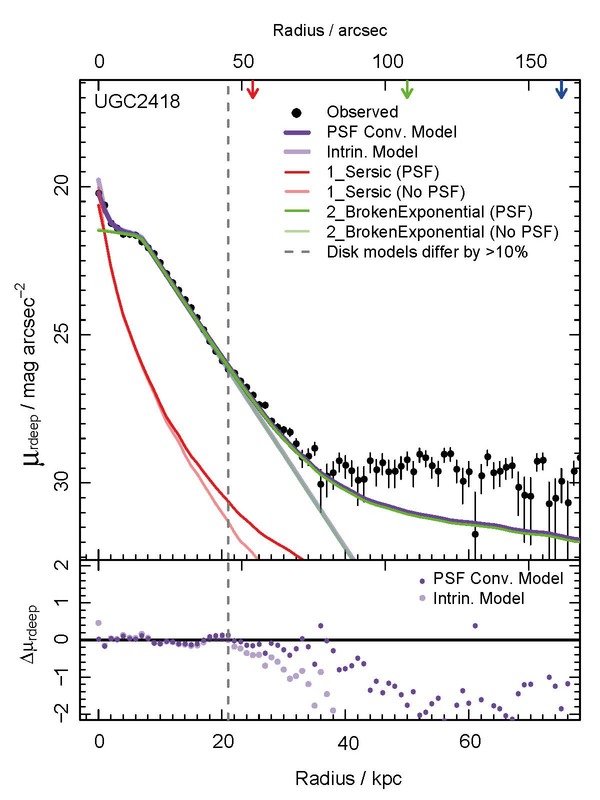}
\caption[Effect of the PSF on the profiles]{Test whether the PSF can create an artificial halo. The red lines are a S\'ersic fit to the inner galaxy, corresponding to the bulge, the light line before and the dark line after convolution with the PSF. The green lines a broken exponential, again before and after convolution. The purple lines correspond similarly the sum of the two. The grey line indicates the minimum radius where the models differ by more than 10\% from the observations. We have IC\,1515 on the left, NGC\,493 in the middle, and UGC\,2418 at the right.}\label{fig:PSF}
\end{figure*}

\begin{figure*}
\centering
\includegraphics[width=0.83\textwidth]{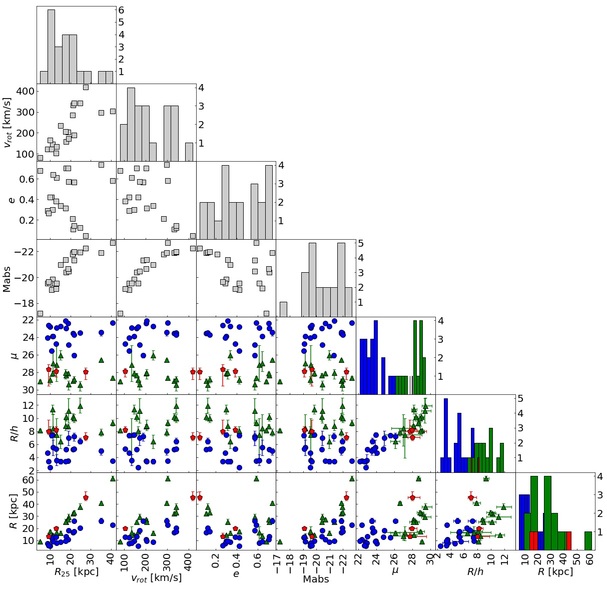}
\caption[Overview of the correlations between various parameters]{Overview of the correlations between various parameters. The absolute magnitude is in the $r'$-band. The brightness of the feature $\mu$ is in $r'$-mag/arcsec$^2$. Where multiple features are shown together the haloes are represented by green triangles, the breaks with blue circles and the truncations with red pentagons.}\label{fig:correlations}
\end{figure*}

As argued by \citet{deJong2008} and \citet{sandin14,sandin15}, the point spread 
function (PSF) in SDSS images can lead to a significant contribution of 
scattered light in the outer regions of galaxies.
This is in particular true for highly inclined objects, where the minor 
axis is much shorter than the major axis, and light from the inner parts 
of a galaxy can thus be scattered to larger projected radii more easily.
We are typically reaching the 29-30 $r'$-mag/arcsec$^2$ depth in our profiles. 
What is the effect of the point spread function at these depths on the 
profiles? 
To test this, we will generate synthetic galaxies based on the fitted 
scale lengths of IC\,1515, NGC\,493 and UGC\,2418.
These galaxies have been selected based on their size and inclination; 
together they represent a cross-section of the various types of galaxies 
in this sample.

For this one needs to know the PSF to very faint levels. A first study was
performed in \cite{peters14} with a carefully build PSF using well over 
10000 stars observed in various $g'$, $r'$ and $i'$ averaged images with 
the \textsc{SExtractor} and \textsc{PSFEx} tools. This gave the PSF up to a 
radius of 177''. The PSF images were convolved with a synthetic image using 
the \textsc{astropy.convolution.convolve} package 
for the convolution, which is a non-FFT convolution algorithm
\citep{Astropy}.
Then  \textsc{ellipse} was used on on both the original and convolved 
synthetic image to extract their profiles. The result of that was
presented as Figure 7.5 in \cite{peters14} and the 
conclusion from that was that the PSF can seriously affect the
outer surface brightness profile, but in general is insufficient to be the only 
cause of the apparent extended stellar halo. 

In the mean time a better PSF was determined using bright stars in addition to a
very large number of faint stars \citep{ft13}. With this a new analysis of the
outer parts of galaxies and the reality of the apparent extended stellar 
haloes is possible \citep{Kelvin15}. Here we report the result of such an 
analysis on the same three galaxies as in \cite{peters14}, using a 
$2001 \times 2001$ PSF. 

First, automated routines to perform image cutting, PSF-resizing, source 
detection, secondary source masking and initial parameter estimation are 
employed for each galaxy via methods similar to those presented in 
\citep{Kelvinea2012}. These data products are subsequently used to fit a 2D 
model to each source using the Imfit software \citep{Erwin2015}. 
A Sersic + Broken
 Exponential model is fit using a smaller $255 \times 255$ 
pixel PSF, for speed. 
Loosely speaking, these components can be interpreted as bulge and disc, 
respectively. The automated fitting outputs are saved, and used as an input 
for a manual fit with the larger $2001 \times 2001$ pixel PSF. 
 Modifications are made to the model parameters to 
provide a more accurate estimation of the underlying galaxy flux. In order 
to facilitate this step, a plotting script generates a series of 2D images 
of the original galaxy, the model, and the sub-components, as well as a measure 
of the 1D surface brightness profile. 

The final
results are shown in Figure \ref{fig:PSF}. The conclusion is the same as 
before. The PSF contamination contribution, defined  as 
[flux(convolved) - flux(intrinsic)] / flux(intrinsic), are 49\%\ for NGC\,493, 
45\%\ for UGC\,2428 and 47\%\ for IC\, 1515. 
It is thus clear that a significant amount of light in the halo is due to the 
PSF, as predicted by \citet{deJong2008} and \citet{sandin14}.
In some cases, such as NGC\,493,  this apparent sterlight-like halo 
could be almost solely due to the PSF effect, while in others it could 
account for only a part the apparent stellar haloes.

\subsection{Correlations}\label{sec:psfdiscussions}
We have performed a wide range of  tests on the various parameters associated 
with the galaxies to look for possible correlations. 
An overview of all the correlations, or lack thereof, is presented in 
Figure \ref{fig:correlations}.
We have found no correlation of either the radius of the feature or its 
brightness at that radius, with morphological type, apparent or absolute 
magnitude, radial velocity, distance, inclination, or maximum rotational 
velocity (see Table \ref{tbl:sample}).

Our main, surprising, result is shown in Figure \ref{fig:kpcs}, where we 
show the radius $R$ of the various types of feature against the radius at 
the 25th magnitude, $R_{25}$, of its host galaxy.
A couple of interesting things can be noticed from this figure. 
First, there is a clear correlation of the break radius $R_b$ with the size 
of the galaxy, with a best linear fit of $R_b = 0.77R_{25}$.
A second, different correlation is that of the radii of the halos \textit{and} 
truncations, $R_h$ and $R_t$, with host galaxy size.
Both features follows the combined linear relation $R_{h\&t}=1.43R_{25}$ (see 
Figure \ref{fig:kpcs} for a larger version of this plot).
Measured separately, the linear fits become $R_h = 1.42R_{25}$ and 
$R_t=1.52R_{25}$.
Table \ref{tbl:main} illustrates this relation between halos and truncations 
further. Halos and truncations are mutually exclusive: a galaxy has either 
a halo or a truncation.

\begin{figure}
\centering
\includegraphics[width=0.25\textwidth]{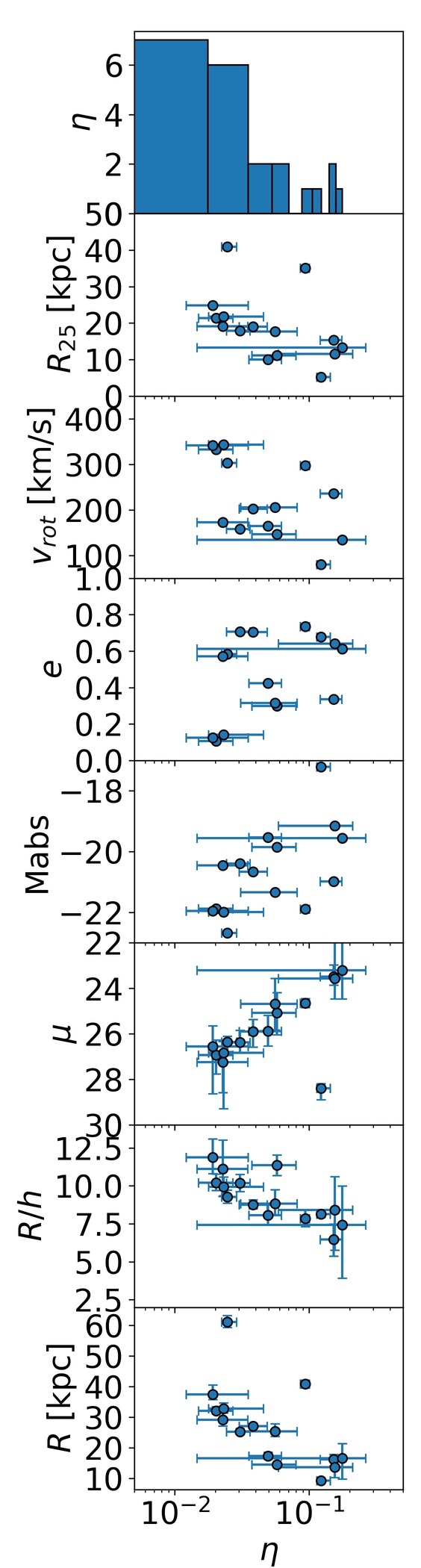}
\caption[Overview of the correlations with the halo light fraction]{Overview of the correlations with the fraction of the total light from the galaxy that is emitted by the halo $\eta$. The absolute magnitude $M_{\rm abs}$ and the brightness of the halo $\mu$ are both in $r'$-mag/arcsec$^2$. The maximum rotation of the galaxy  $V_{\rm rot}$  is in km/s. The ratio of the profile scale length $h$ to the radius of the halo $R/h$ is dimensionless.}\label{fig:correlationshalo}
\end{figure}

\begin{figure}
 \centering
  \includegraphics[width=0.48\textwidth]{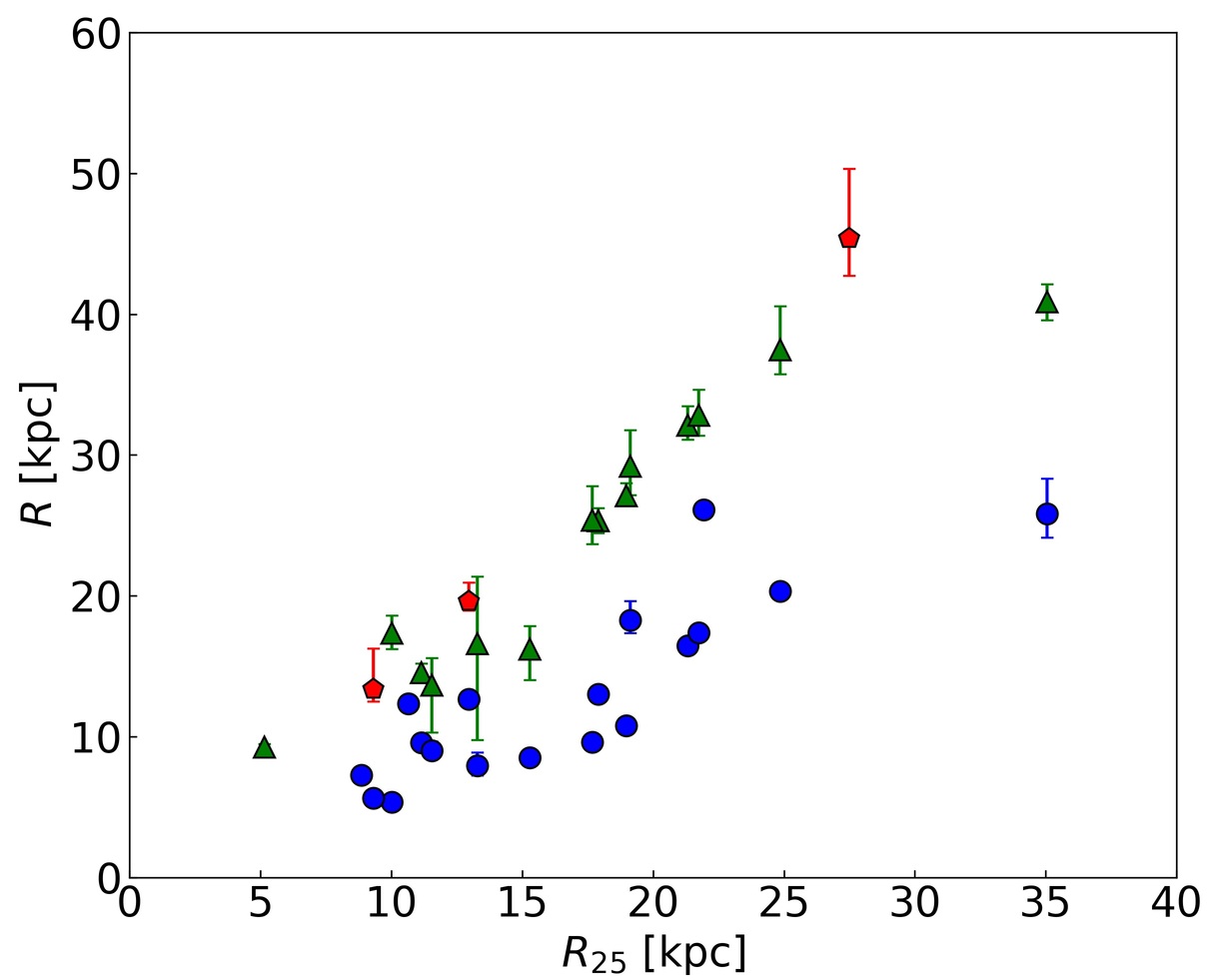}
  \caption[Correlation of $R_{25}$ with various feature radii]{The correlation of the $R_{25}$, indicative of the size of a galaxy, with the various feature radii $R$. Red pentagons represent the truncations. Blue circles represent the breaks. Green triangles represent the haloes. }
  \label{fig:kpcs}
\end{figure}

In Figure \ref{fig:correlationshalo} we present an overview of the 
correlation of the halo light fraction $\eta$ with various other parameters 
of the galaxy.
The halo light fraction correlates most strongly with the position at which 
the halo starts to dominate the profile, expressed as $R/h$, and the 
brightness of the halo at that position $\mu$. These are unsurprising 
correlations, as a bright halo can be expected to dominate a profile at 
higher luminosities than a faint halo.
More interesting is the lack of correlation with ellipticity $e$ and 
maximum circular rotation $V_{\rm rot}$ of the galaxy.
The relative brightness of the halo does not depend on inclination nor 
dynamical mass.
The only other correlations are between the size $R_{25}$ and absolute 
brightness $M_{\rm abs}$ 
of the galaxy, and the halo fraction. Fainter and smaller 
galaxies have the relatively largest haloes.

\subsection{Breaks}

We postulate that the breaks we see well within the discs of our sample 
galaxies occur in the general radial region where the star-forming spiral 
arms end. This can be recognized rather clearly in individual galaxies when 
comparing the radii of the breaks with the images of the galaxies, as shown 
in the appendix figures.

A similar distinction between breaks and truncations has been reported 
by \citet{mbt12} from their study of highly inclined galaxies. The results 
are not comparable in detail, because it is not clear how the end of a 
star-forming spiral arm zone is observed in an inclined galaxy, where 
dust obscuration and projection effects will hinder direct observations. 
Nevertheless, the typical  radii and surface brightness levels for the breaks 
found by \citet{mbt12}, of $7.9$\,kpc and $22.5$\,mag\,arcsec$^{-2}$, 
respectively, are compatible with the values we find here for face-on 
galaxies, $13.6\pm6.4$\,kpc and $22.8\pm2.8$\,$r'$-mag\,arcsec$^{-2}$.
Selecting all variations of type II profiles in \citet[Table 3]{pt06}, 
we find an average of $9.5\pm3.9$\,kpc, which is again compatible.

We conclude that breaks at a level of 
around $23$\,$r'$-mag/arcsec$^{2}$ are very common in disc galaxies, and 
that in all cases there is a steeper slope after the break. Such breaks 
are most likely due to changes in the star formation properties. These 
occur as one changes from the regime of actively star-forming spiral arms 
to more quiescent parts of the disc. They may also be related to the star 
formation often occurring near the ends of a bar, which can lead to a rise 
in the radial profile, followed by a break as the radius moves outside the 
bar regime. Note that evidence for a break should not come from the rectified 
polar plots, but only from the ellipse fitting or the EPs.

In any case, we conclude that these disc breaks are unrelated to the 
truncations which we discuss elsewhere in this paper, and which we 
claim are the counterparts to the truncations reported since decades 
in edge-on galaxies. Claims in the literature of truncations (as well 
as anti-truncations) in face-on galaxies at or near the relatively high 
surface brightness levels identified here with disc breaks should not 
be confused or identified with truncations, nor should they be referred 
to as such.

\subsection{Truncations}
The criteria for the classification of the features in the outer parts 
at fainter levels than breaks are rather unambiguous. In order to test 
the presence of a truncation we exclusively examined the rectified polar 
plots and only accept eveidence for a truncation when a change in slope 
(down-bending) occurs at levels below 26 {\muu} in at least one of the RPP the 
profiles. Any truncation should show up best when teh isophote used for the 
rectfication is close to the surface brightness where the truncation starts. 
Only in the case of NGC\,799 do we see the profiles continue without 
any trace of a change in slope. In most cases there either is a clear 
downturn of the radial luminosity profile --and then we identify that as 
a truncation-- or the profile flattens off and fades away into the noise 
quite slowly --the signature of a faint halo. The existence of a truncation 
need not be confirmed by requiring that it also occur in the other profiles. 
After all, the rectified polar plots were especially designed to 
optimally show truncations, but need not necesserily be absent in 
other methods, or in all other RPP profiles. 
It is true that in the PAS profiles, they should be present also, but in 
practice, those profiles can never be traced out far enough to do this 
confirmatory test. 

In most cases (15 out of the 22 galaxies), there is very clear evidence 
of a change in slope in the profiles towards a slower decline with 
galactocentric radius at faint levels. This we identify with faint 
apparently stellar haloes. By the time the luminosity profiles of these haloes 
can be identified the surface brightness has fallen from 26.5 to 28 
$r'$-mag/arcsec$^{2}$ (see Section \ref{sec:stellarhaloes}). Among the 
remaining seven cases, there are four galaxies in which there is no 
indication for a faint stellar halo and where the luminosity 
profiles  show no evidence for a downturn. Therefore, in these galaxies 
we have no indication for a truncation. In the remaining three cases, we 
have been able to identify truncations. Inspection of the polar plots 
suggests that the slopes and the surface brightness of the changes are 
such that if these systems were seen edge-on, they would show the same 
signature of a truncation as we indeed observe in real edge-ons.
It may be of interest to comment on the similarity of the profiles of 
UGC\,12208 and 12709 to the double-downbending profiles in \citet{pt06} and 
\citet{Herr13}, in both of which one galaxies display such a feature. In 
NGC\,4517A the second break occurs at a much brighter level and in DDO\,125 is 
a much fainter dwarf galaxy than we have in our sample. So, we do not believe 
that this a phenomenon similar to our truncations.

\begin{figure*}
\centering
\includegraphics[width=0.37\textwidth]{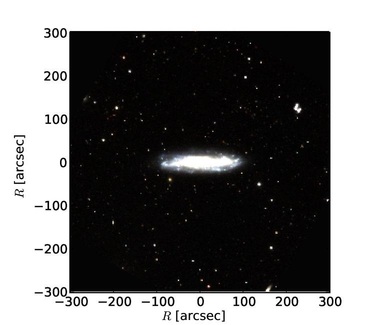}
\includegraphics[width=0.37\textwidth]{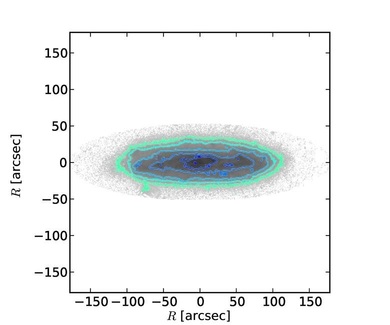}
\includegraphics[width=0.37\textwidth]{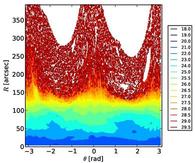}
\includegraphics[width=0.37\textwidth]{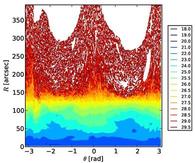}
\caption[PAS plots foe the highly inclined galaxy NGC\,493.]{Polar plots for the highly inclined galaxy NGC\,493. Note that the fainter levels of surface brightness extend further out along the minor axis at position angles 0 and $\pi$ radians, even after rectification.}  \label{fig:PPRhalo}
\end{figure*}

\begin{figure*}
\centering
\includegraphics[width=0.67\textwidth]{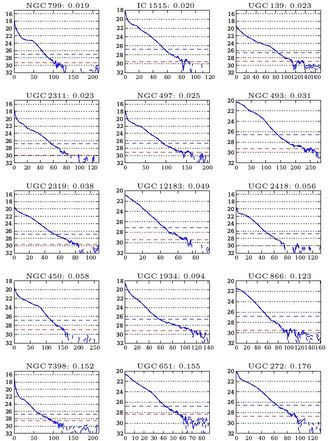}
\caption[Profile mosaic depicting the increase of the halo light fraction across the sample]{Profile mosaic depicting the increase of the halo light fraction $\eta$ across the sample, 
the value of which is shown below each figure. 
The surface brightness $\mu$ in $r'$-mag/arcsec$^2$ on the vertical axis. The radius $R$ in arcsec on the bottom horizontal axis.} \label{fig:eta1}
\end{figure*}

\begin{figure*}
\centering
\includegraphics[width=0.67\textwidth]{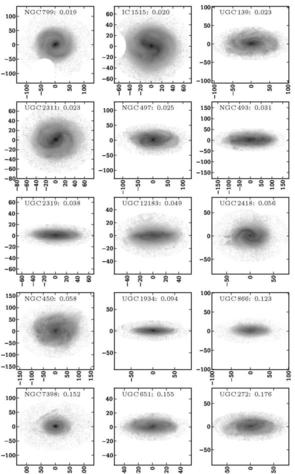}
\caption[Image mosaic depicting the increase of the halo light fraction across the sample]{Image mosaic depicting the increase of the halo light fraction $\eta$ across the sample, the value of which is shown in each figure. The axis scales are in arcsec.}  \label{fig:eta2}
\end{figure*}

The truncations reported here occur at radii of significantly more
than the canonical $4.2\pm0.5$ scale lengths, reported by \citet{vdks81a,vdks82}
in edge-on galaxies, and certainly the  $3.6\pm0.6$ of \citet{kk04}. 
The best estimate of the face-on surface brightness
at which truncations observed in edge-on systems in the \citet{kk04} sample
should become evident, would be of order  $25.3\pm0.6$ R-mag arcsec$^{-2}$.
Consistent with this, \citet{vanderKruit2008A} showed that the radii
of the `classical truncations' of \citet{pt06} systematically occur
at smaller radii than edge-on truncations in galaxies with the same 
rotation velocity. The truncations found in this paper thus occur at a 
larger number of scalelenghts 
and at fainter surface brightnesses than in edge-on galaxies. 
This also holds when we use other samples of edge-on galaxies such as 
\citet{Pohlen2000}. We have no real explanation for this; 
the only argument we can offer is that there may
be serious problems comparing face-on and edge-on scalelengths and inferring
face-on surface brightness levels from photometry in edge-on galaxies.
The obvious difference is that in 
edge-ons the fitting is done above and below the dust lane and therefore 
excludes effects young populations and dust. It requires a separate 
investigation to address this fundamental issue.

The most straightforward interpretation would be that most or almost all 
galaxies have truncations of the sort we have found in three systems in 
this paper, but that their presence in inclined or face-on 
views is hidden by the presence of  faint outer stellar haloes, in the way 
proposed by \citet{bt13} and \citet{mtk14}. This
halo may to some extent at least actually be due to PSF scattered 
light. In this case, the galaxy itself is hiding the presence of the 
truncation. On the other hand, if the absence of a detectable halo is 
unrelated to the presence or absence of a truncation then we are forced 
to conclude that three out of seven discs are truncated.  
For edge-on systems, \citet{kkg02} and \citet{kk04} 
found from a sample of 34 southern, edge-on spiral galaxies that at least 20 
of these have radial truncations.  The rest  either have no truncations, or 
have them at a level inaccessible for us now. We have vindicated the 
prediction by \citet{bt13} that (spiral) galaxies usually --as 
judged from our sample, two-thirds of all galaxies-- have faint stellar 
haloes that prevent the identification of a truncation.

\subsection{Haloes}\label{sec:stellarhaloes}
A natural prediction of the collapsing galaxy formation scenarioes, in
particular the hierarchical one, is the 
ubiquitous presence of stellar haloes surrounding galaxies 
\citep[e.g.][]{els62,sz78, bc01,sg03,sm95}. These extended and 
diffuse stellar components are formed from the debris of disrupted 
satellites accreted along the cosmic time \citep[e.g.][]{bj05,bkg03,ans06}.

In the local Universe, both based on integrated photometry 
\citep[e.g.][]{bt13} and on star counts 
\citep[e.g.][]{Barker2012}, the outer regions of low-inclined disc 
galaxies ($R > 10$\,kpc) are often characterized by an exponential surface 
brightness decrement followed by an excess of light over this exponential 
decay. 
This  outer excess of light is located at $R \geq 20$\,kpc and has surface 
brightness $\mu_R > 28$\,mag\,arcsec$^{-2}$. 
These faint surface brightness levels are equivalent to the ones found in 
the literature using galaxies with edge-on orientations at similar radial 
distances above the galactic planes. 
For instance, the edge-on spiral NGC\,4565 has, along its minor axis, a 
surface brightness of $\mu_{6660} = 27.5$\,mag\,arcsec$^{-2}$ 
(i.e., $\mu_R \sim 28$\,mag\,arcsec$^2$ in the R-band) at 22 kpc above the 
disc \citep{Wu2002}. 
Similarly, \citet{Jablonka2010} found that the stellar halo of the edge-on 
disc NGC\,3957 has a surface brightness of 28.5 mag arcsec$^{-2}$ in the 
R-band at 20 kpc above the disc plane.
The similarity between these values in edge-on and face-on galaxies would 
entice us to conclude that we are observing the same component of the galaxy, 
i.e., their stellar haloes.

A property of a stellar halo (and for that matter for the distribution
of scattered light by the PSF) would be that it more or less shperical, 
independent of the inclination of the galaxy. It is the case that we fixed the 
ellipticity of the isophotes in the ellipse-fitting procedure, so we
have no information on this in the outcome of these fits. 
In particualr, in adge-on galaxies, the outer light should be more spherical 
than the galaxy. This should show up as faint extensions of the light 
distribution in the polar plots, even after rectification. And it should be 
most visible in highly inclined systems as extensions of the surface 
brightnesses at levels fainter than 27 or 28 mag arcsec$^{-2}$ in the RPP 
profiles near the minor axis (position angles 0 and $\pi$ radians. 

The two galaxies in Fig.~\ref{fig:PolarPlots} are too face-on to demonstrate 
this effect. But NGC\,493 should serve as a good example, see 
Fig.~\ref{fig:PPRhalo}. The fainter surface brightnesses indeed extend 
further out along the minor axis, indicating the the faint light has 
a more or less spherical distribution but stretched in the production
of the polar plots, as
expected for a stellar halo (but also for the case of scattered light).
We see similar behaviour for the other highly inclined galaxies in our sample, 
NGC\,497, 1090, UGC\,139, 272, 651, 866, 1934, 12183, 1218 and 32319 (see
the figures in the online Appendix). To test this assertion further, we have 
attempted to perform ellipse fits to several of our galaxies, in which we let 
the ellipticity run free into the very outskirts. The aim was to demonstrate 
that the ellipticity increases towards these radii. Unfortunately, we find 
that the ellipse task fails to converge at the outskirts of galaxies.

We have demonstrated in Section \ref{sec:psfdiscussion} that for the 
three galaxies tested, the PSF scattered light of the galaxies was 
responsible for most or in one case even all of the observed halo. 
Further support for the importance of PSF scattering comes from a comparison
of the fraction of light $\eta$ in the `haloes' and the total dynamical mass 
of the galaxies. It turns out that the two are essentially uncorrelated.
This is at odds with numerical simulations, which predict that the fraction of 
mass in  stellar haloes should increase with  total stellar mass 
\citep{Cooper2013}. 
As argued in Section \ref{sec:psfdiscussion}, the smallest and faintest 
galaxies have the highest halo light fractions $\eta$.
As an aid to the reader, in Figures \ref{fig:eta1} and \ref{fig:eta2} we 
demonstrate in two mosaics how the halo light fraction $\eta$ increases in 
the sample.

Interestingly, our observed halo-light fractions (0.02-0.03) are already 
below the theoretical values of \citet{Cooper2013} ($\sim0.1$; their Fig. 
12 for B/T$<$0.2). However, there is room for explaining this discrepancy. 
First, our data could be selected against finding massive haloes. In fact, 
our sample of spiral galaxies is biased towards objects which do not have 
signatures of distortions caused by on-going or recent mergers. Second, we 
have studied the fraction of light of the haloes in the $r'$-band instead of 
the fraction of stellar mass, as done in the simulations. If, as expected, 
the stellar populations in the haloes are older  than those in the galaxy 
discs, then our observed light fractions in the $r'$-band will be smaller 
than the stellar mass fractions. A rough estimate of this bias can be reached 
as follows. Assuming that the mean luminosity weighted age of the stellar 
population in the discs is ~1-2 Gyr and in the stellar haloes ~6-8 Gyr, then 
assuming a Solar metallicity and a Kroupa IMF, according to the 
\citet{Vazdekis2012} models we would get $M/L_{r'}$\,(disc)=0.6-1.0 and 
$M/L_{r'}$\,(stellar halo)=2.4-2.8. Consequently,  our data (at least the 
observed offset between the fraction of light in the haloes and the 
theoretical expectation for the fraction of mass) could be in agreement 
with the numerical expectations.

Our result should serve as a warning for any future work: the 
effect of the PSF must be taken into account when attempting to interpret 
surface photometry of stellar haloes in face-on galaxies.
The PSF will always play a major, limiting role in the detections.
In this sense, a better and safer method for finding stellar haloes would be 
detecting individual stars, such as done by \citet{Radburn-Smith2011A}.

\section*{Conclusions}\label{sec:conclusions}
We present surface photometry of 22 nearby face-on and moderately inclined 
galaxies, from the IAC Stripe82 Legacy Project, using (co-)added data from the 
$g'$, $r'$ and $i'$ bands. 
Using  traditional ellipse profiles, as well as our new rectified polar 
profiles, this allows us to probe down to a surface brightness level of 
29-30 $r'$-magnitude / arcsec$^2$.
Due to this, ellipse profiles smooth out any truncation that may occur and 
a special technique, introduced here and named rectified polar plots, is 
vital to detect them.
For the first time ever, we have been able to detect truncations in three 
of our 22 face-on galaxies, using rectified polar profiles.
Furthermore, we find that 15 other galaxies have apparent stellar haloes.
Using synthetic galaxies convolved with a real PSF, we exclude the possibility 
that these  haloes could be solely explained due to scattered light 
from the PSF.
The light scattered by the PSF is, however, the dominant source of light 
in the outer parts of these galaxies.
The presence of haloes and truncations is mutually exclusive, and we argue that 
the presence of a stellar halo and/or light scattered light by the PSF
often outshines any truncation that might be present.

The radius of the onset of  truncations and haloes correlates tightly with the 
galaxy size, as measured with the $R_{25}$ parameter.
We find that the correlation is effectively the same for both features.
We have also detected 17 breaks, which are found much closer to the centre of 
galaxies.
Breaks are also found to correlate with the size of the galaxy, again 
measured with $R_{25}$.
We have found no correlation between the colour or radius at the onset 
of breaks, truncations and stellar haloes, with morphological type, 
apparent or absolute magnitude, radial velocity and distance, 
inclination and maximum rotational velocity.

\section*{Acknowledgments}
SPCP is grateful to the Space Telescope Science Institute, Baltimore, USA, the 
Research School for Astronomy and Astrophysics, Australian National University, 
Canberra, Australia, and the Instituto de Astrofisica de Canarias, La Laguna, 
Tenerife, Spain, for hospitality and support during  short and extended
working visits in the course of his PhD thesis research. He thanks
Roelof de Jong and Ron Allen for help and support during an earlier 
period as visiting student at Johns Hopkins University and 
the Physics and Astronomy Department, Krieger School of Arts and Sciences 
for this appointment.

PCK thanks the Instituto de Astrof\'{\i}sica de Canarias, La Laguna, Tenerife 
for hospitality during two visits supported by the Severo Ochoa mobility 
program, during which most of this work was performed.
PCK is also grateful to the directors of the Space Telescope Science Institute 
in Baltimore, USA and the Research School of Astronomy and Astrophysics, Mount 
Stromlo Observatory, Australian National University at Canberra, Australia
for hospitality during numerous work visits, of which a number were directly 
related to this research. He also acknowledges his local hosts
Ron Allen, Ken Freeman and Johan Knapen for help and support.

We are grateful to Chris Sandin and Roelof de Jong for stressing the issue of the effects of the Point Spread Function.

We acknowledge financial support to the DAGAL network from the People Programme 
(Marie Curie Actions) of the European Unionbs Seventh Framework Programme FP7/2007-2013/ 
under REA grant agreement number PITN-GA-2011-289313, and from the Spanish Ministry of 
Economy and Competitiveness (MINECO) under grant numbers AYA2013-48226-C3-1-P (IT) and 
AYA2013-41243-P (JHK). 

Work visits by SPCP and PCK have been supported by an annual grant 
from the Faculty of Mathematics and Natural Sciences of 
the University of Groningen to PCK accompanying of his distinguished Jacobus 
C. Kapteyn professorhip and by the Leids Kerkhoven-Bosscha Fonds. PCK's work
visits were also supported by an annual grant from the Area  of Exact 
Sciences of the Netherlands Organisation for Scientific Research (NWO) in 
compensation for his membership of its Board.


\bibliography{refsVII}
\bibliographystyle{mn2e}



\appendix


\section{Overview of individual galaxies}\label{sec:overview}
The following panels show our results for the individual galaxies.

In the top-left panel all surface brightness profiles are shown. The blue line 
shows the conventional ellipse profile. {\bf To show the uncertainty created 
by noise, we determined the ellipse profiles two additional times. In the 
first, we added the pixel-to-pixel noise to the image, while in the second 
we subtracted this amount. We show these two additional profiles by dashed 
blue profiles.}
Offset by +0.5 mag are the equivalent 
profiles, in green. Three versions of corrected polar profiles 
(using rectification at 25, 26 and 27 magnitudes arcsec$^-2$, respectively) 
are shown in purple with an offset of -0.5 mag. The PAS profiles are 
shown in red and are offset +1 mag from the central value of the ellipse 
profiles. We note that the PAS are in pseudo $r'$-mag/arcsec$^2$ and do not 
represent the same scale as the other profile (see above 
for more details). The dashed blue vertical line denotes the maximum radius 
used for the PAS and EP profile extraction. The dashed horizontal blue line 
represents the $1\sigma$ standard deviation in the pixel-to-pixel noise 
(see section 3). The 
red dashed horizontal line represents the uncertainty limit.
We use the Python  \textsc{BAYES\_MVS} package to establish the 
confidence limits on the background zeropoint. The  \textsc{BAYES\_MVS} 
package provides a 90\%\ confidence range for this zeropoint. We define the 
uncertainty limit as the upper confidence limit of this range. This provides 
a natural limit to the accuracy, below which the uncertainty in the background 
zeropoint is too high for any reliable profile.
The top-right panel shows the sky in counts as seen from the elliptical 
profiles.
In the second row we show on the left side a RGB image produced from the 
$g'$, $r'$ and $i'$ images, rotated such that the major axis of the galaxy 
aligns with the horizontal axis. The scale for all images is in arcsec 
unless otherwise noted. On the right side we show the uncorrected background 
of the (co-)added image, smoothed by a Gaussian with a radius of three pixels. 
The image is again aligned to the major axis of the galaxy. The colours range 
from -0.5 to 0.5 ADU. The green contour represents a outer cutout region from 
the larger raw image. The ticks beside each image are in arcsec.
The left panel on the third row shows the corrected background in the cutout 
region. The scale again ranges from -0.5 to 0.5 ADU. No smoothing is applied. 
The blank holes are due to the mask. The blue ellipse shows the ellipse used 
as a limit for the PAS and EP methods.
The right panel on the third row shows the cutout ellipse of the PAS and EP 
methods. The scale has been optimized to show the faint outer regions of the 
galaxy. The contours match the colourbars on the last row.
The last row shows two versions of the polar profiles, with on the left the 
uncorrected profile, and on the right the effect of a rectification at a 
level of 26 magnitudes / arcsec$^2$.

\begin{figure*}
 \centering 
\includegraphics[width=0.37\textwidth]{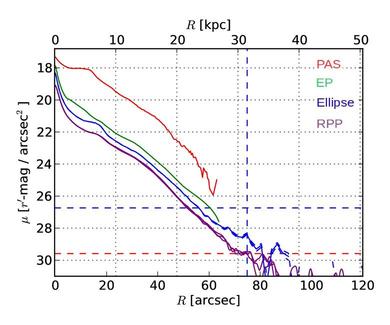}
\includegraphics[width=0.37\textwidth]{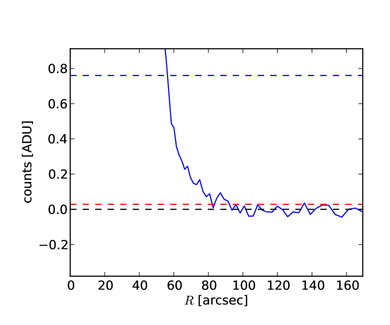}
\includegraphics[width=0.37\textwidth]{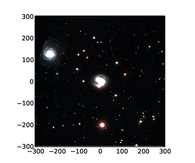}
\includegraphics[width=0.37\textwidth]{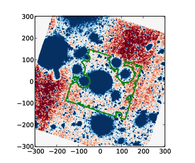}
\includegraphics[width=0.37\textwidth]{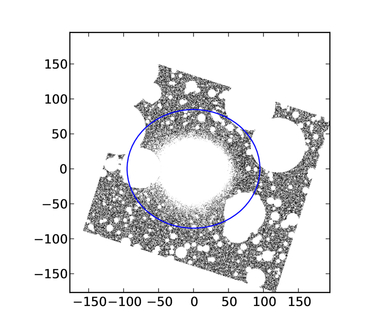}
\includegraphics[width=0.37\textwidth]{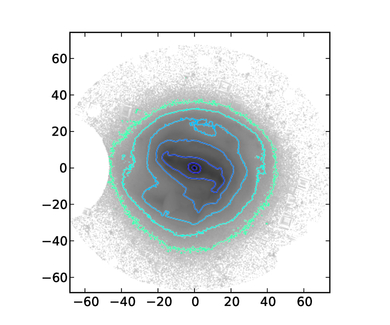}
\includegraphics[width=0.37\textwidth]{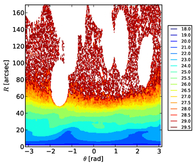}
\includegraphics[width=0.37\textwidth]{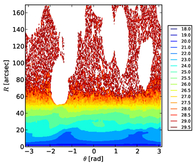}
\caption{Results for IC\,1515.}\label{fig:IC1515}
\end{figure*}

\begin{figure*}
 \centering 
\includegraphics[width=0.37\textwidth]{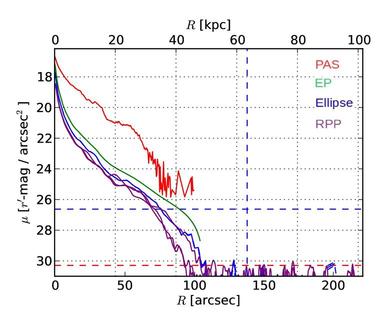}
\includegraphics[width=0.37\textwidth]{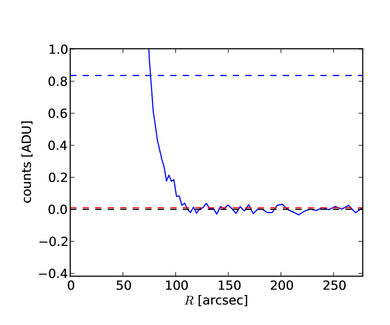}
\includegraphics[width=0.37\textwidth]{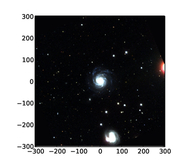}
\includegraphics[width=0.37\textwidth]{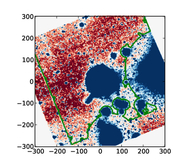}
\includegraphics[width=0.37\textwidth]{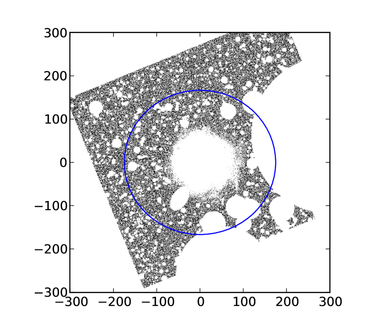}
\includegraphics[width=0.37\textwidth]{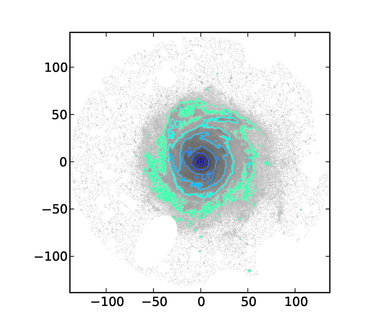}
\includegraphics[width=0.37\textwidth]{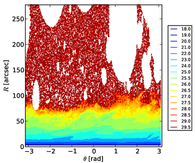}
\includegraphics[width=0.37\textwidth]{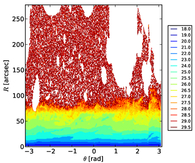}
\caption{Results for IC\,1516.}\label{fig:IC1516}
\end{figure*}

\begin{figure*}
 \centering 
\includegraphics[width=0.37\textwidth]{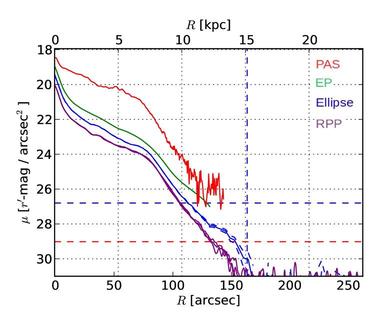}
\includegraphics[width=0.37\textwidth]{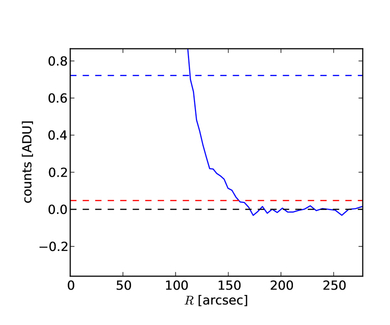}
\includegraphics[width=0.37\textwidth]{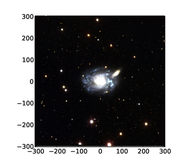}
\includegraphics[width=0.37\textwidth]{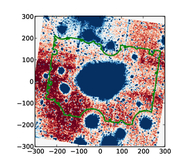}
\includegraphics[width=0.37\textwidth]{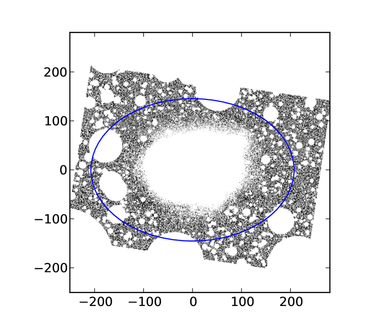}
\includegraphics[width=0.37\textwidth]{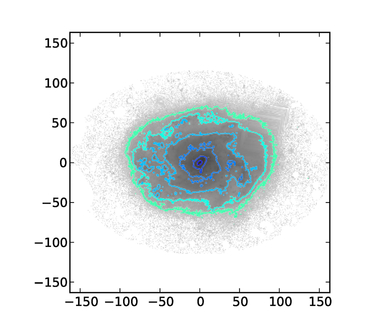}
\includegraphics[width=0.37\textwidth]{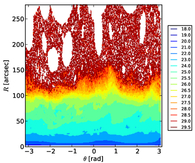}
\includegraphics[width=0.37\textwidth]{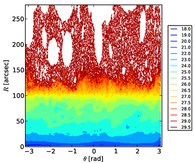}
\caption{Results for NGC\,450.}\label{fig:NGC450}
\end{figure*}

\begin{figure*}
 \centering 
\includegraphics[width=0.37\textwidth]{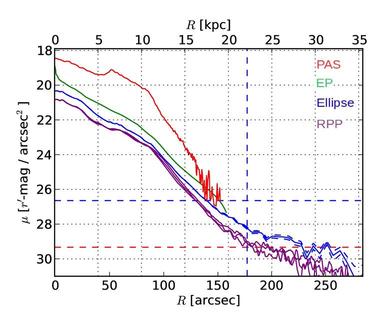}
\includegraphics[width=0.37\textwidth]{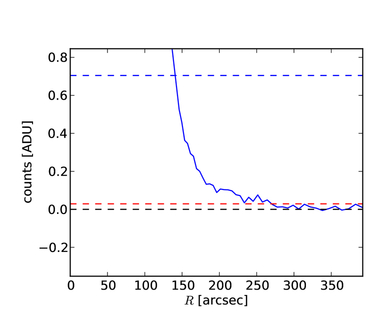}
\includegraphics[width=0.37\textwidth]{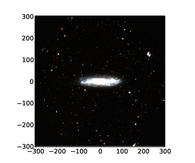}
\includegraphics[width=0.37\textwidth]{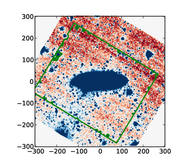}
\includegraphics[width=0.37\textwidth]{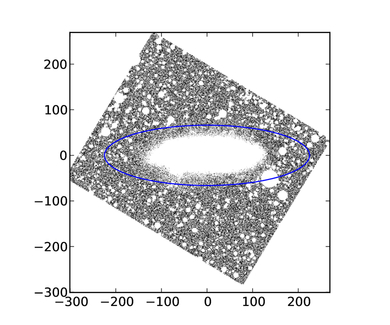}
\includegraphics[width=0.37\textwidth]{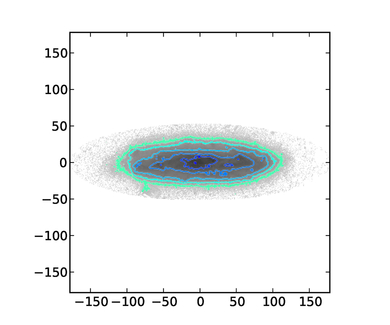}
\includegraphics[width=0.37\textwidth]{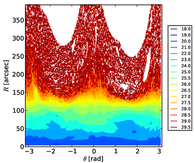}
\includegraphics[width=0.37\textwidth]{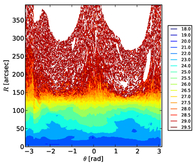}
\caption{Results for NGC\,493.}\label{fig:NGC493}
\end{figure*}

\begin{figure*}
 \centering 
\includegraphics[width=0.37\textwidth]{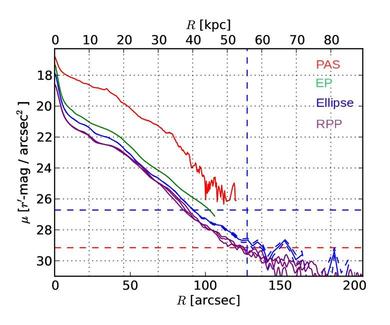}
\includegraphics[width=0.37\textwidth]{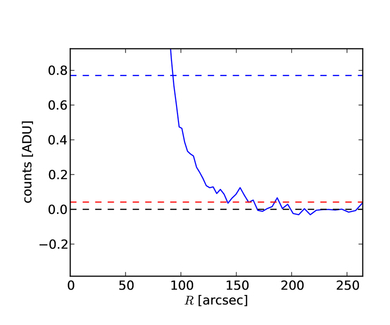}
\includegraphics[width=0.37\textwidth]{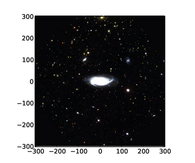}
\includegraphics[width=0.37\textwidth]{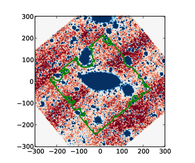}
\includegraphics[width=0.37\textwidth]{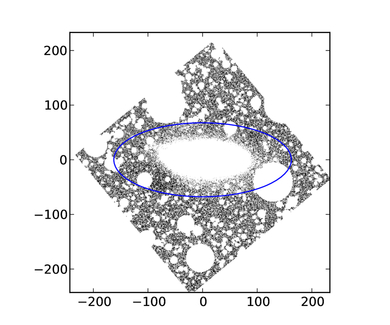}
\includegraphics[width=0.37\textwidth]{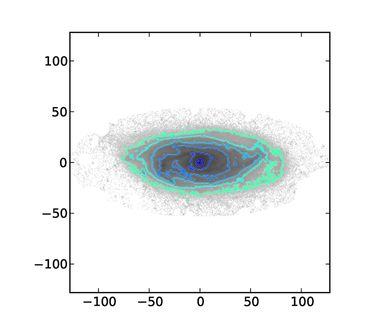}
\includegraphics[width=0.37\textwidth]{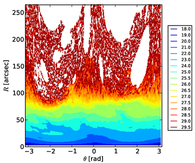}
\includegraphics[width=0.37\textwidth]{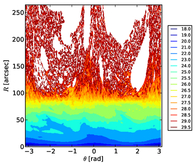}
\caption{Results for NGC\,497.}\label{fig:NGC497}
\end{figure*}

\begin{figure*}
 \centering 
\includegraphics[width=0.37\textwidth]{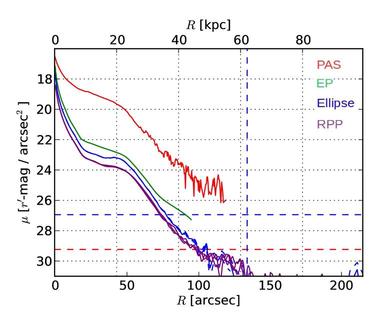}
\includegraphics[width=0.37\textwidth]{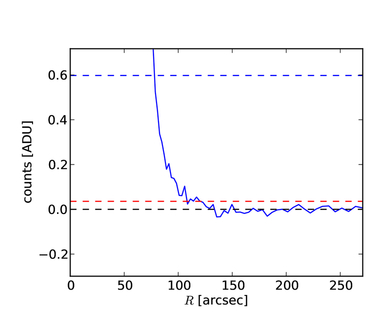}
\includegraphics[width=0.37\textwidth]{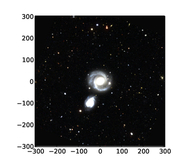}
\includegraphics[width=0.37\textwidth]{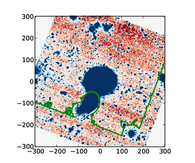}
\includegraphics[width=0.37\textwidth]{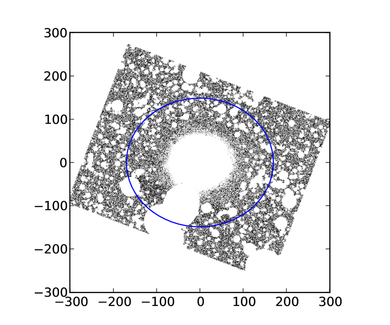}
\includegraphics[width=0.37\textwidth]{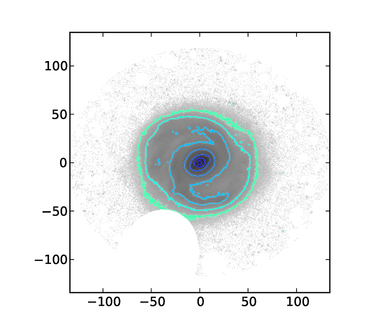}
\includegraphics[width=0.37\textwidth]{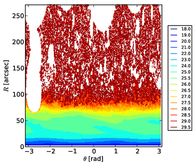}
\includegraphics[width=0.37\textwidth]{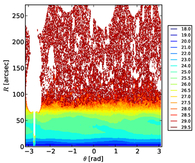}
\caption{Results for NGC\,799.}\label{fig:NGC799}
\end{figure*}

\begin{figure*}
 \centering 
\includegraphics[width=0.37\textwidth]{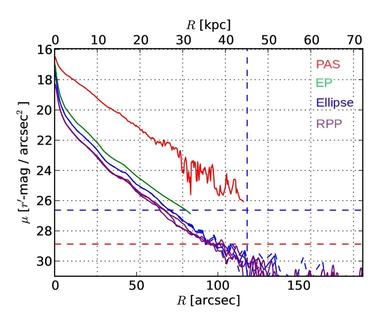}
\includegraphics[width=0.37\textwidth]{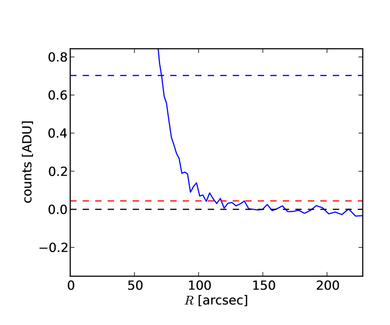}
\includegraphics[width=0.37\textwidth]{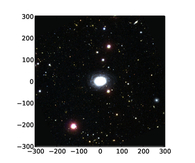}
\includegraphics[width=0.37\textwidth]{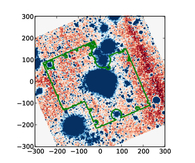}
\includegraphics[width=0.37\textwidth]{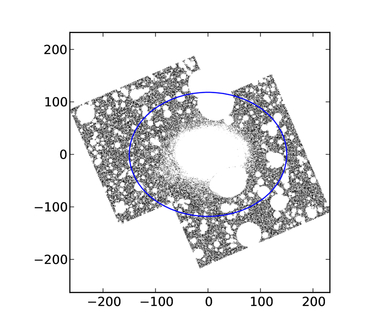}
\includegraphics[width=0.37\textwidth]{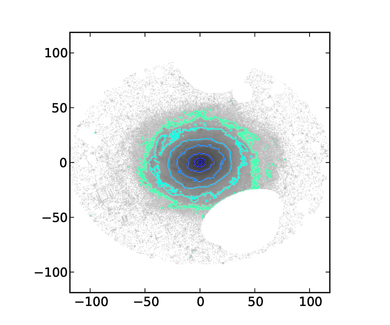}
\includegraphics[width=0.37\textwidth]{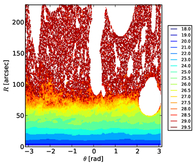}
\includegraphics[width=0.37\textwidth]{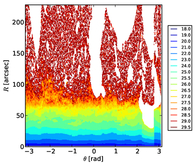}
\caption{Results for NGC\,856.}\label{fig:NGC856}
\end{figure*}

\begin{figure*}
 \centering 
\includegraphics[width=0.37\textwidth]{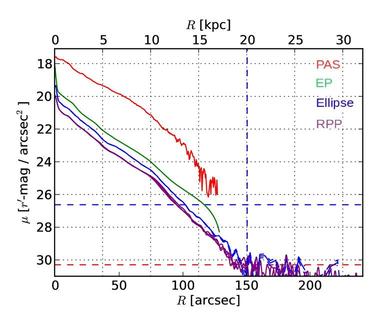}
\includegraphics[width=0.37\textwidth]{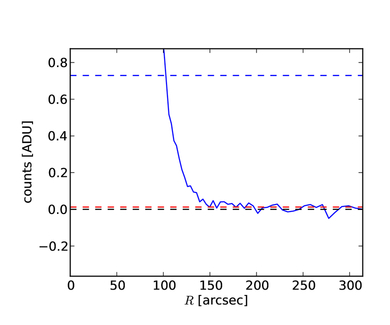}
\includegraphics[width=0.37\textwidth]{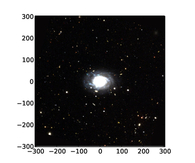}
\includegraphics[width=0.37\textwidth]{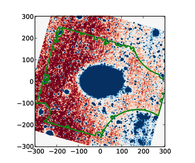}
\includegraphics[width=0.37\textwidth]{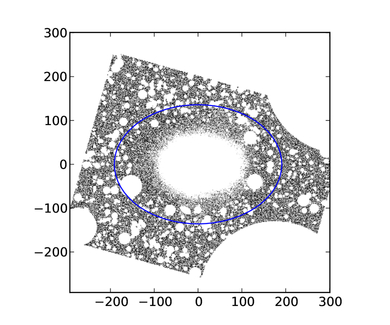}
\includegraphics[width=0.37\textwidth]{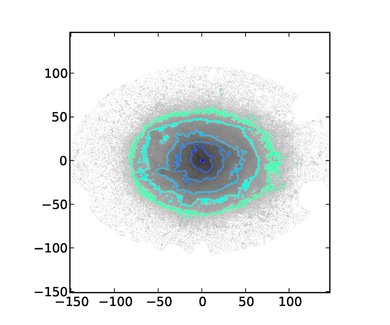}
\includegraphics[width=0.37\textwidth]{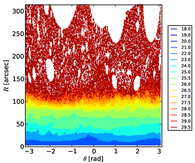}
\includegraphics[width=0.37\textwidth]{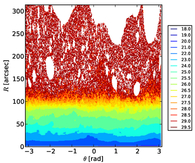}
\caption{Results for NGC\,941.}\label{fig:NGC941}
\end{figure*}

\begin{figure*}
 \centering 
\includegraphics[width=0.37\textwidth]{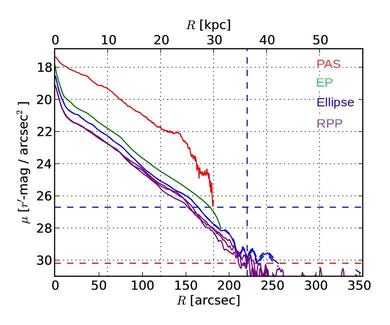}
\includegraphics[width=0.37\textwidth]{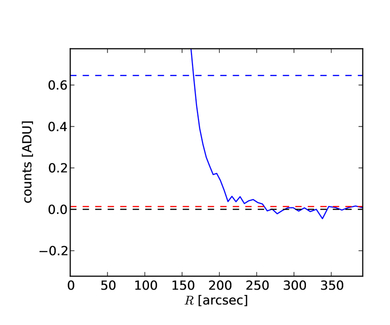}
\includegraphics[width=0.37\textwidth]{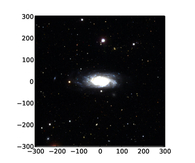}
\includegraphics[width=0.37\textwidth]{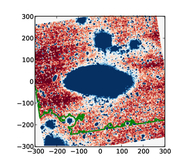}
\includegraphics[width=0.37\textwidth]{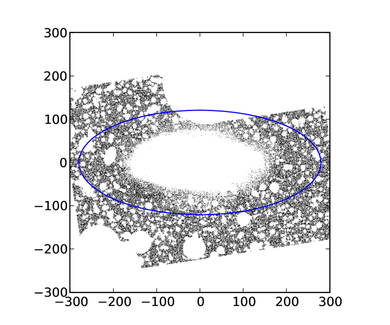}
\includegraphics[width=0.37\textwidth]{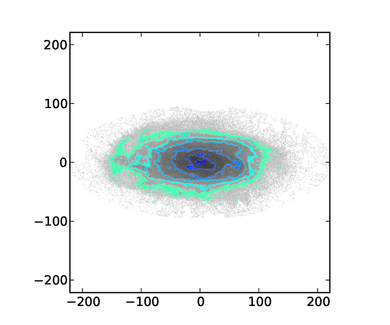}
\includegraphics[width=0.37\textwidth]{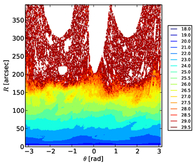}
\includegraphics[width=0.37\textwidth]{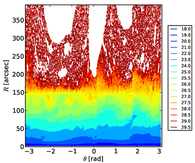}
\caption{Results for NGC\,1090.}\label{fig:NGC1090}
\end{figure*}

\begin{figure*}
 \centering 
\includegraphics[width=0.37\textwidth]{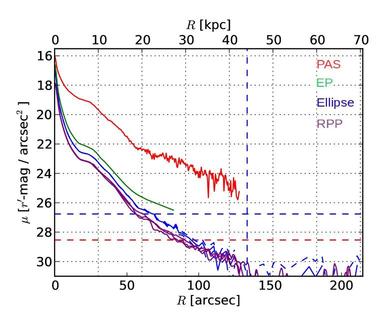}
\includegraphics[width=0.37\textwidth]{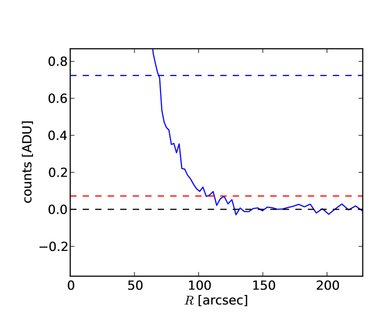}
\includegraphics[width=0.37\textwidth]{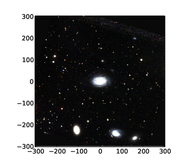}
\includegraphics[width=0.37\textwidth]{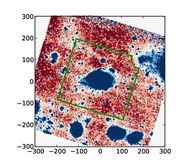}
\includegraphics[width=0.37\textwidth]{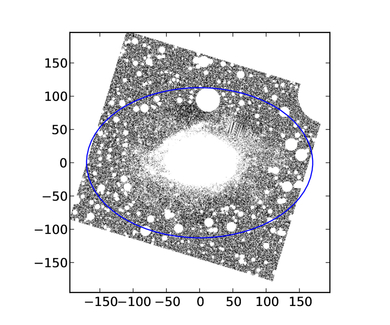}
\includegraphics[width=0.37\textwidth]{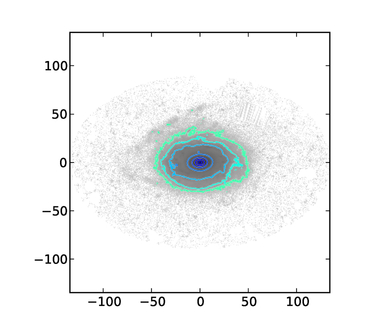}
\includegraphics[width=0.37\textwidth]{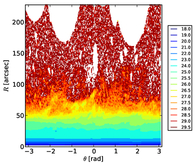}
\includegraphics[width=0.37\textwidth]{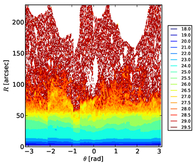}
\caption{Results for NGC\,7398.}\label{fig:NGC7398}
\end{figure*}

\begin{figure*}
 \centering 
\includegraphics[width=0.37\textwidth]{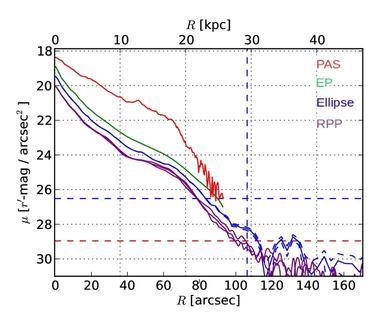}
\includegraphics[width=0.37\textwidth]{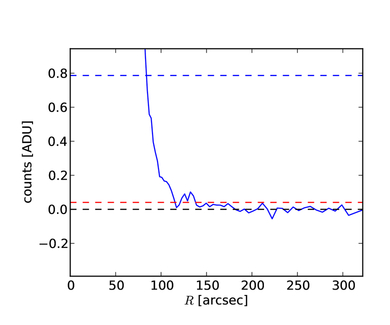}
\includegraphics[width=0.37\textwidth]{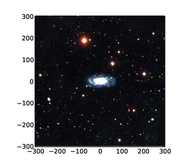}
\includegraphics[width=0.37\textwidth]{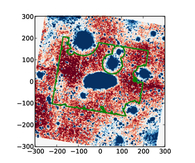}
\includegraphics[width=0.37\textwidth]{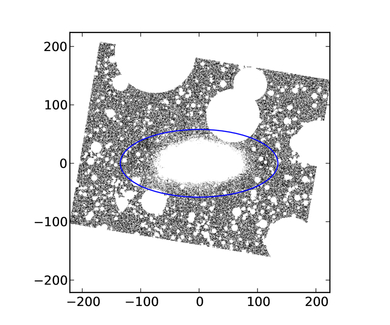}
\includegraphics[width=0.37\textwidth]{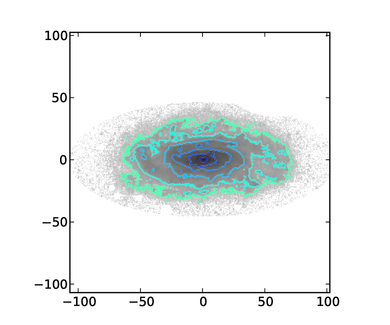}
\includegraphics[width=0.37\textwidth]{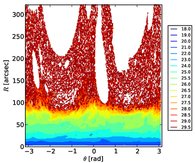}
\includegraphics[width=0.37\textwidth]{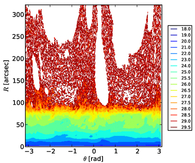}
\caption{Results for UGC\,139.}\label{fig:UGC139}
\end{figure*}

\begin{figure*}
 \centering 
\includegraphics[width=0.37\textwidth]{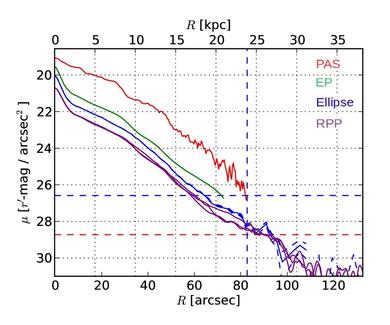}
\includegraphics[width=0.37\textwidth]{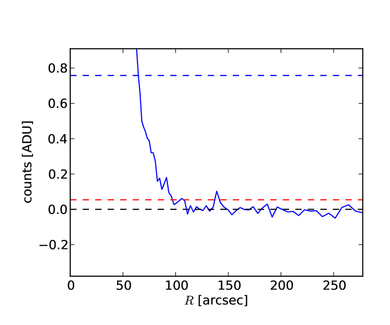}
\includegraphics[width=0.37\textwidth]{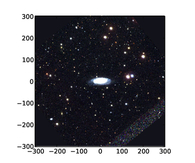}
\includegraphics[width=0.37\textwidth]{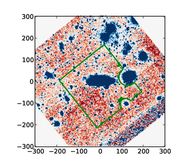}
\includegraphics[width=0.37\textwidth]{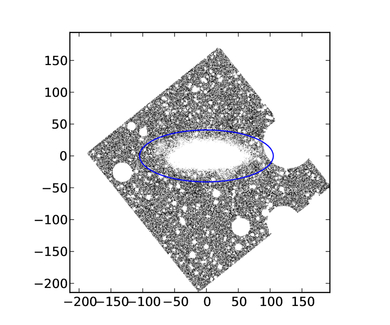}
\includegraphics[width=0.37\textwidth]{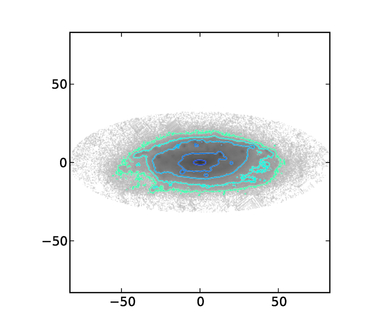}
\includegraphics[width=0.37\textwidth]{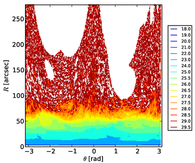}
\includegraphics[width=0.37\textwidth]{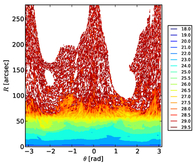}
\caption{Results for UGC\,272.}\label{fig:UGC272}
\end{figure*}

\begin{figure*}
 \centering 
\includegraphics[width=0.37\textwidth]{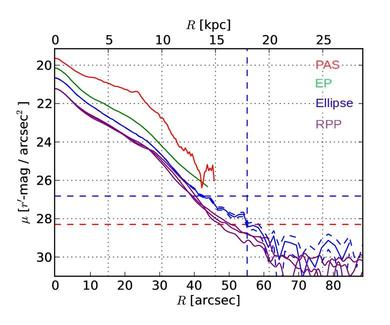}
\includegraphics[width=0.37\textwidth]{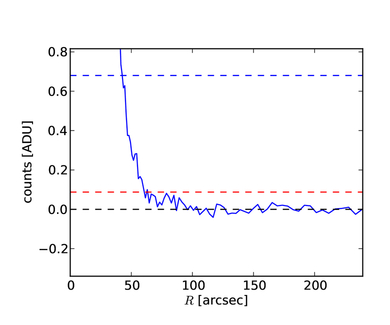}
\includegraphics[width=0.37\textwidth]{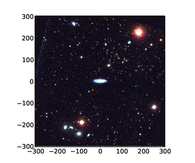}
\includegraphics[width=0.37\textwidth]{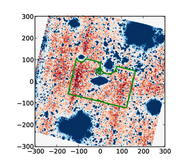}
\includegraphics[width=0.37\textwidth]{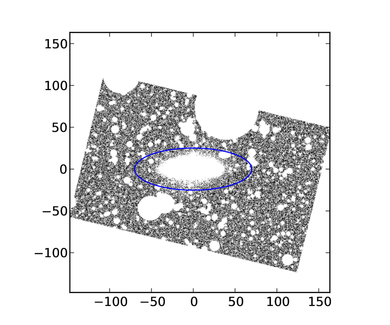}
\includegraphics[width=0.37\textwidth]{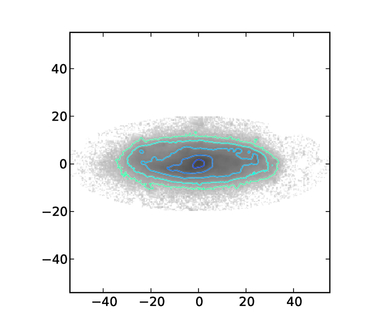}
\includegraphics[width=0.37\textwidth]{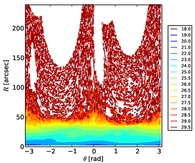}
\includegraphics[width=0.37\textwidth]{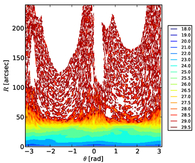}
\caption{Results for UGC\,651.}\label{fig:UGC651}
\end{figure*}

\begin{figure*}
 \centering 
\includegraphics[width=0.37\textwidth]{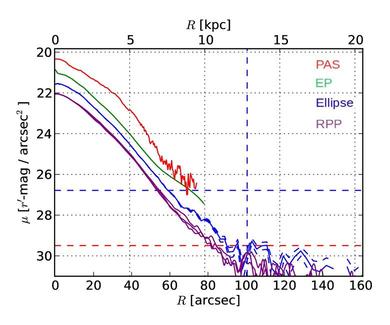}
\includegraphics[width=0.37\textwidth]{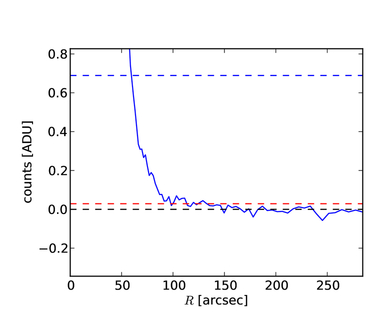}
\includegraphics[width=0.37\textwidth]{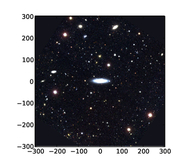}
\includegraphics[width=0.37\textwidth]{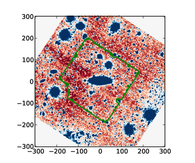}
\includegraphics[width=0.37\textwidth]{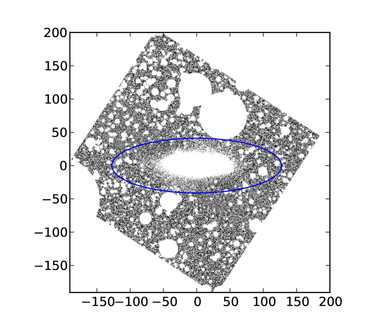}
\includegraphics[width=0.37\textwidth]{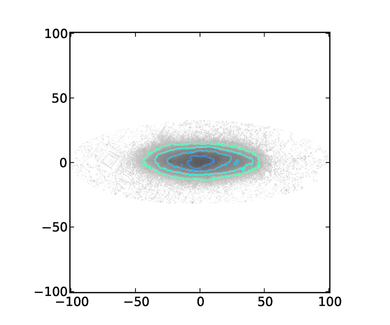}
\includegraphics[width=0.37\textwidth]{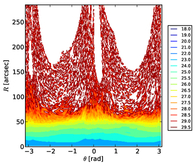}
\includegraphics[width=0.37\textwidth]{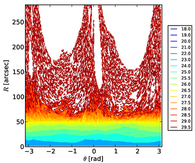}
\caption{Results for UGC\,866.}\label{fig:UGC866}
\end{figure*}

\begin{figure*}
 \centering 
\includegraphics[width=0.37\textwidth]{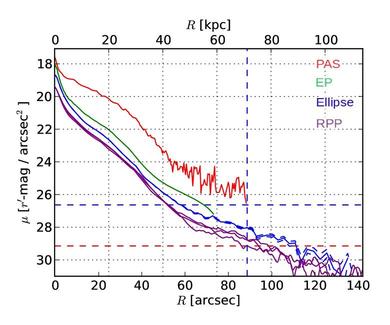}
\includegraphics[width=0.37\textwidth]{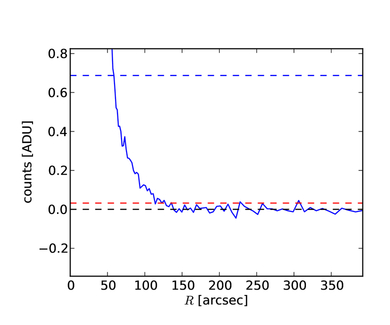}
\includegraphics[width=0.37\textwidth]{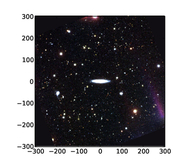}
\includegraphics[width=0.37\textwidth]{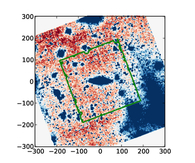}
\includegraphics[width=0.37\textwidth]{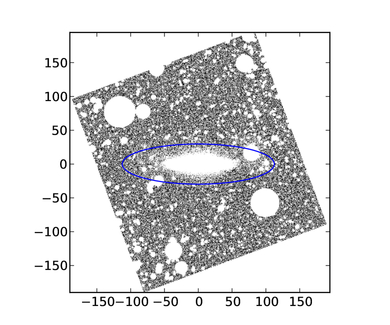}
\includegraphics[width=0.37\textwidth]{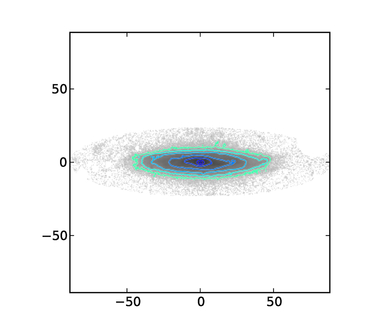}
\includegraphics[width=0.37\textwidth]{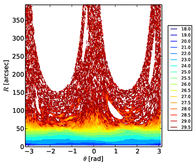}
\includegraphics[width=0.37\textwidth]{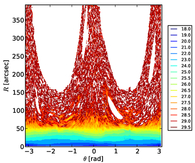}
\caption{Results for UGC\,1934.}\label{fig:UGC1934}
\end{figure*}

\begin{figure*}
 \centering 
\includegraphics[width=0.37\textwidth]{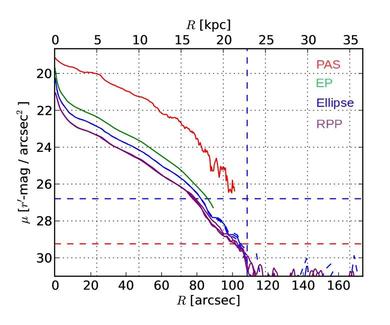}
\includegraphics[width=0.37\textwidth]{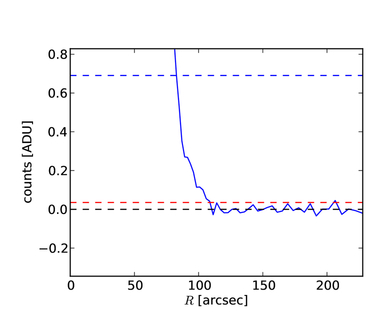}
\includegraphics[width=0.37\textwidth]{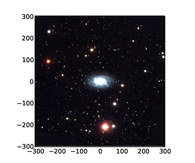}
\includegraphics[width=0.37\textwidth]{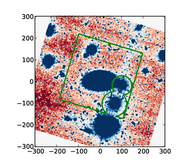}
\includegraphics[width=0.37\textwidth]{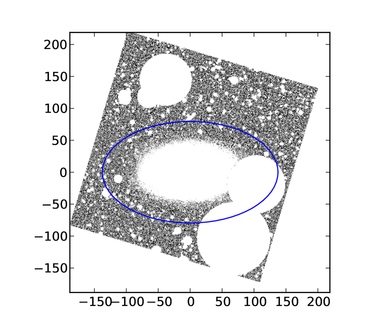}
\includegraphics[width=0.37\textwidth]{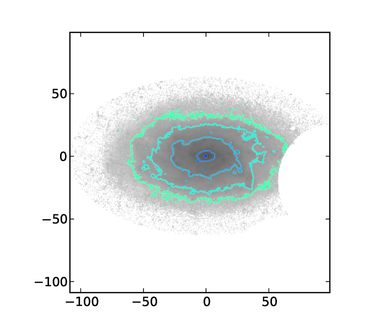}
\includegraphics[width=0.37\textwidth]{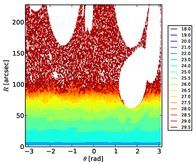}
\includegraphics[width=0.37\textwidth]{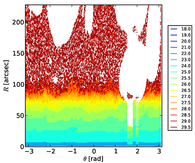}
\caption{Results for UGC\,2081.}\label{fig:UGC2081}
\end{figure*}
\clearpage

\begin{figure*}
 \centering 
\includegraphics[width=0.37\textwidth]{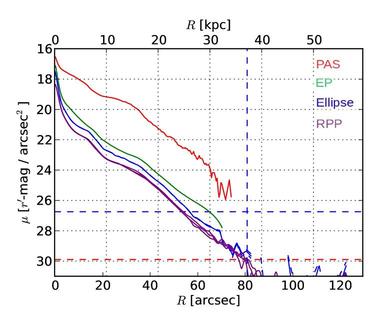}
\includegraphics[width=0.37\textwidth]{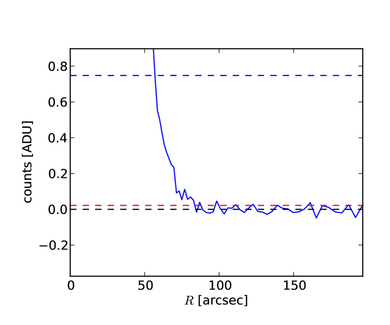}
\includegraphics[width=0.37\textwidth]{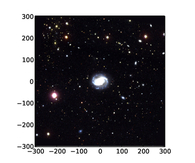}
\includegraphics[width=0.37\textwidth]{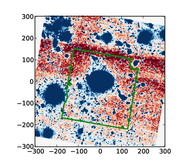}
\includegraphics[width=0.37\textwidth]{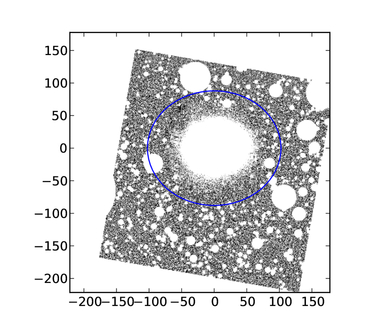}
\includegraphics[width=0.37\textwidth]{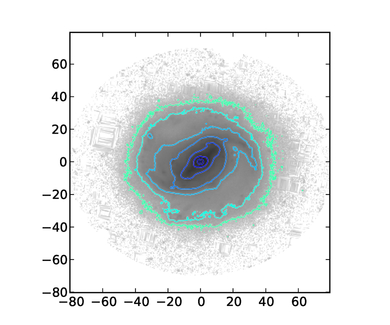}
\includegraphics[width=0.37\textwidth]{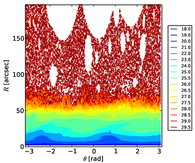}
\includegraphics[width=0.37\textwidth]{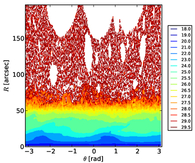}
\caption{Results for UGC\,2311.}\label{fig:UGC2311}
\end{figure*}

\begin{figure*}
 \centering 
\includegraphics[width=0.37\textwidth]{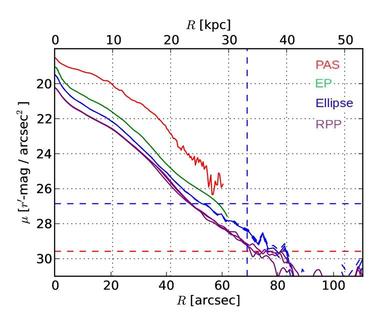}
\includegraphics[width=0.37\textwidth]{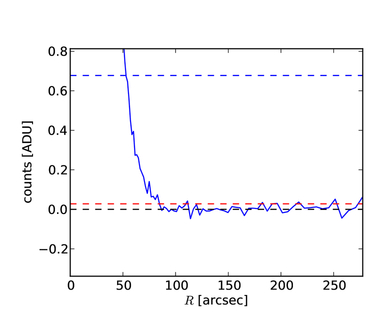}
\includegraphics[width=0.37\textwidth]{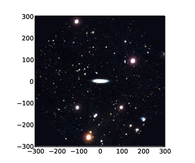}
\includegraphics[width=0.37\textwidth]{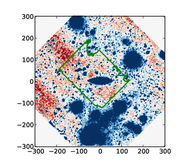}
\includegraphics[width=0.37\textwidth]{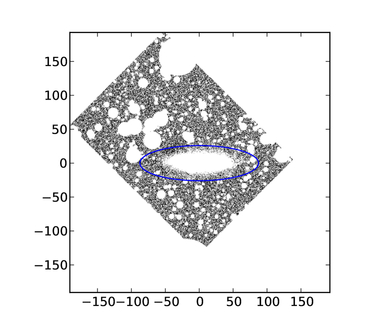}
\includegraphics[width=0.37\textwidth]{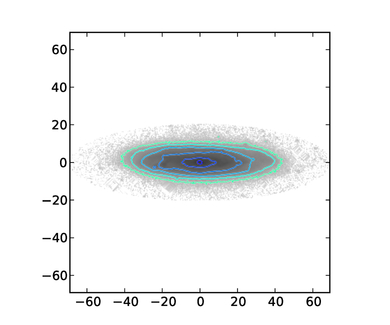}
\includegraphics[width=0.37\textwidth]{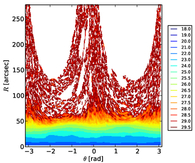}
\includegraphics[width=0.37\textwidth]{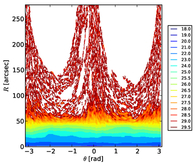}
\caption{Results for UGC\,2319.}\label{fig:UGC2319}
\end{figure*}

\twocolumn
\clearpage

\begin{figure*}
 \centering 
\includegraphics[width=0.37\textwidth]{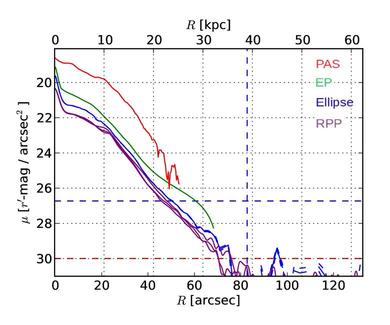}
\includegraphics[width=0.37\textwidth]{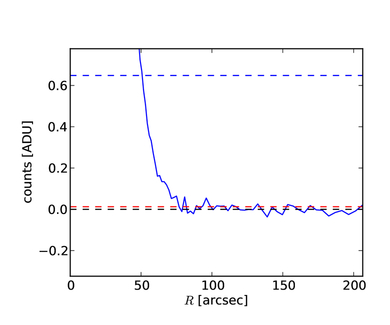}
\includegraphics[width=0.37\textwidth]{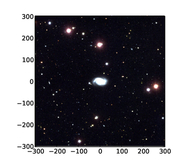}
\includegraphics[width=0.37\textwidth]{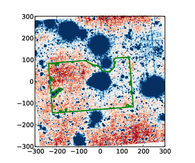}
\includegraphics[width=0.37\textwidth]{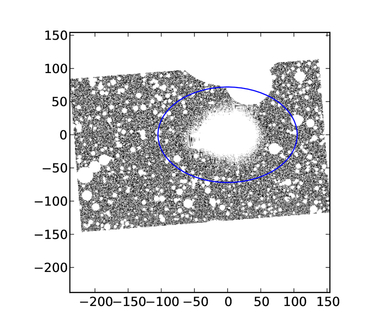}
\includegraphics[width=0.37\textwidth]{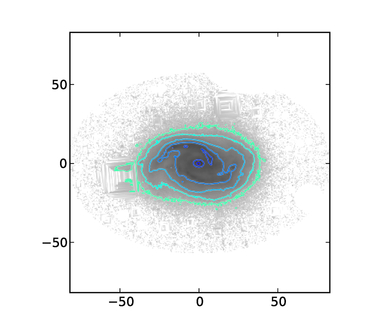}
\includegraphics[width=0.37\textwidth]{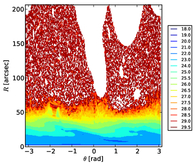}
\includegraphics[width=0.37\textwidth]{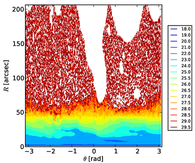}
\caption{Results for UGC\,2418.}\label{fig:UGC2418}
\end{figure*}

\begin{figure*}
 \centering 
\includegraphics[width=0.37\textwidth]{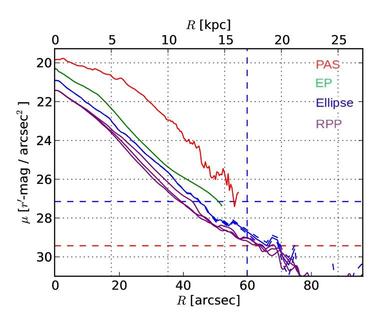}
\includegraphics[width=0.37\textwidth]{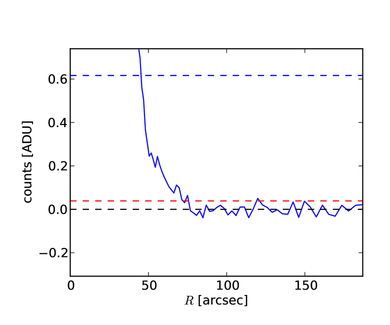}
\includegraphics[width=0.37\textwidth]{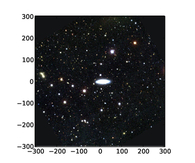}
\includegraphics[width=0.37\textwidth]{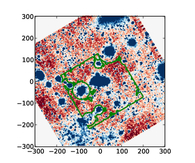}
\includegraphics[width=0.37\textwidth]{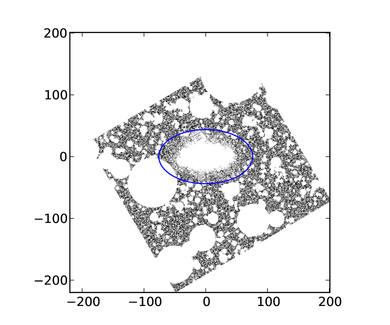}
\includegraphics[width=0.37\textwidth]{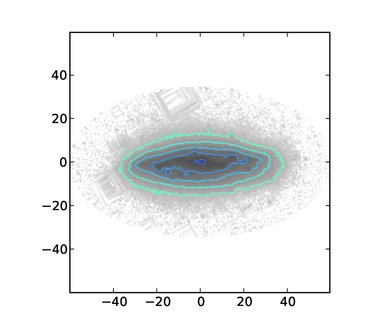}
\includegraphics[width=0.37\textwidth]{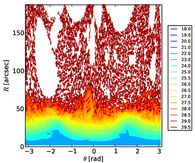}
\includegraphics[width=0.37\textwidth]{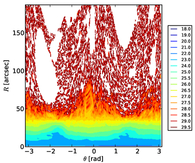}
\caption{Results for UGC\,12183.}\label{fig:UGC12183}
\end{figure*}

\begin{figure*}
 \centering 
\includegraphics[width=0.37\textwidth]{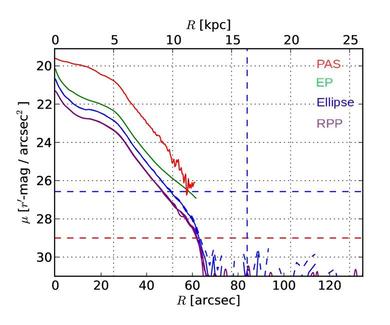}
\includegraphics[width=0.37\textwidth]{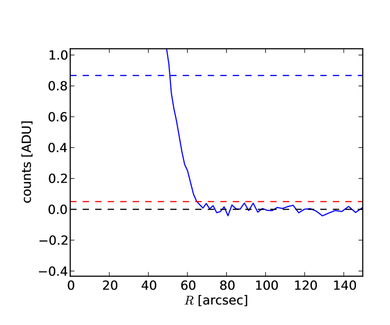}
\includegraphics[width=0.37\textwidth]{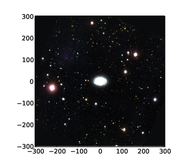}
\includegraphics[width=0.37\textwidth]{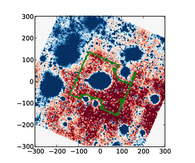}
\includegraphics[width=0.37\textwidth]{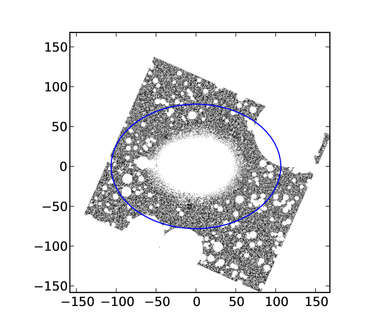}
\includegraphics[width=0.37\textwidth]{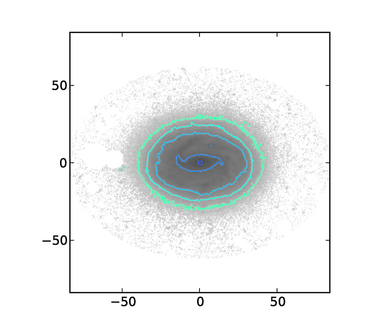}
\includegraphics[width=0.37\textwidth]{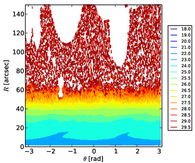}
\includegraphics[width=0.37\textwidth]{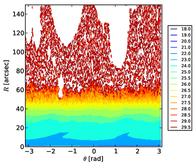}
\caption{Results for UGC\,12208.}\label{fig:UGC12208}
\end{figure*}

\begin{figure*}
 \centering 
\includegraphics[width=0.37\textwidth]{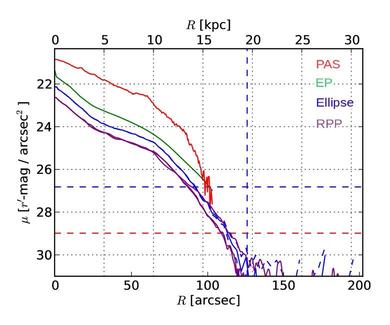}
\includegraphics[width=0.37\textwidth]{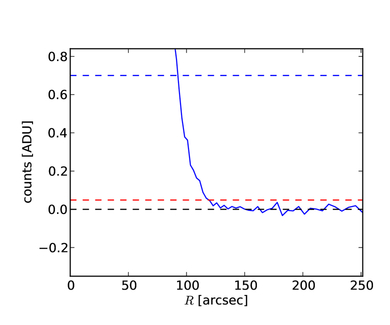}
\includegraphics[width=0.37\textwidth]{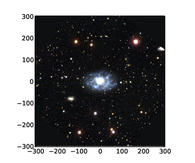}
\includegraphics[width=0.37\textwidth]{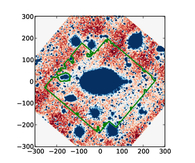}
\includegraphics[width=0.37\textwidth]{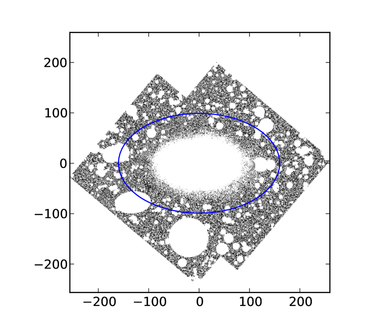}
\includegraphics[width=0.37\textwidth]{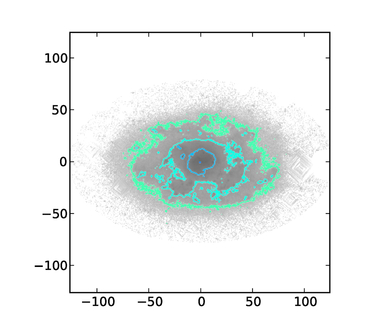}
\includegraphics[width=0.37\textwidth]{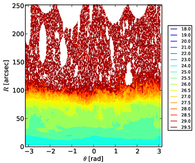}
\includegraphics[width=0.37\textwidth]{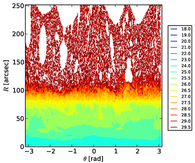}
\caption{Results for UGC\,12709.}\label{fig:UGC12709}
\end{figure*}

\begin{figure*}
\centering
\includegraphics[width=1\textwidth]{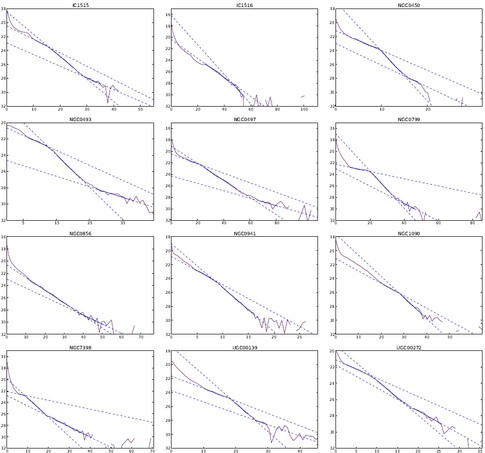}
\caption{{\bf Results of the fits to the elliptical profiles. Magnitudes/arcsec$^2$ on the vertical axis, $R$ in kpc on the horizontal axis.}}\label{fig:allfits1}
\end{figure*}

\begin{figure*}
\centering
\includegraphics[width=1\textwidth]{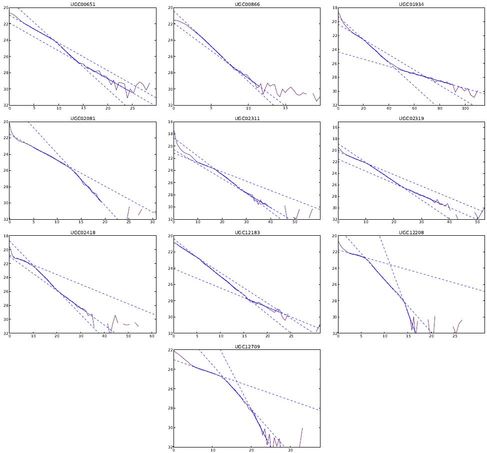}
\caption{{\bf Results of the fits to the elliptical profiles. Magnitudes/arcsec$^2$ on the vertical axis, $R$ in kpc on the horizontal axis.}}\label{fig:allfits2}
\end{figure*}

\twocolumn

\bsp

\label{lastpage}

\end{document}